\documentclass[prd,aps,amssymb,amsmath,tightenlines,nofootinbib]{revtex4}
\usepackage{graphicx,epsfig}
\usepackage{slashed}
\usepackage[colorinlistoftodos]{todonotes}
\usepackage{appendix}

\newcommand{\be}{\begin{equation}}
\newcommand{\ee}{\end{equation}}
\newcommand{\nt}[1]{{\textcolor{black}{#1}}}
\newcommand{\ntt}[1]{{\textcolor{black}{#1}}}

\def\cPT{$\mathcal P \mathcal T~$}
\def\eq{\eqref}

\begin{document}

\preprint[{\leftline{KCL-PH-TH/2021-{\bf 85}}
% \vspace{1cm}

\title{\cPT~symmetric fermionic field theories with axions: \\Renormalisation and dynamical mass generation }

\author{N. E. Mavromatos$^{a, b}$}
\author{Sarben Sarkar$^b$}
\author{A Soto$^{c,d}$}

\affiliation{$^a$Physics Department, School of Applied Mathematical and Physical Sciences, National Technical University of Athens , Athens 157 80, Greece\\
$^b$Theoretical Particle Physics and Cosmology Group, Department of Physics, King's~College~London, Strand, London WC2R 2LS, UK\\
$^c$School of Mathematics, Statistics and Physics, Newcastle University, Newcastle upon Tyne, NE1 7RU, UK\\
$^d$Instituto de Fisica, Pontificia Universidad Catolica de Chile , Santiago 7820436, Chile
}

\begin{abstract}
We consider the renormalisation properties of non-Hermitian Yukawa theories involving a\\ pseudoscalar (axion) field at or near $4$ dimensions. The non-Hermiticity is \cPT-symmetric where $\mathcal P$ is a linear idempotent operator (such as parity) and $\mathcal T$ is an anti-linear idempotent operator (such as time-reversal). The coupling constants of the Yukawa and quartic scalar coupling terms reflect this non-Hermiticity. The path integral representing the field theory is used to discuss the Feynman rules associated with the field theory.  The fixed point structure associated with the renormalisation group has \cPT- symmetric  and Hermitian fixed points. At two loops in the massless theory, we demonstrate the flow from Hermitian to non-Hermitian fixed points. From the one-loop renormalisation of a massive Yukawa theory, a self-consistent Nambu-Jona Lasinio gap equation is established and its  real  solutions are discussed.
 
\end{abstract}

\date{\today}

\maketitle

\section{Introduction} 

Quantum field theories  have provided successful theories of fundamental  (high-energy particle) physics,  of condensed matter and (of aspects) of gravitational physics. In order to go beyond the field-theoretic description provided by the Standard Model (SM) of particle physics, it is necessary to extend the framework.   Two such extensions are:
\begin{itemize}
  \item the first, within the familiar quantum mechanical assumption of Hermiticity, the SM framework is embedded in more general approaches, exemplified by grand unified theories, supersymmetry, supergravity, and string/brane theory (in higher spatial dimensions).
  \item the second, starting in 1998~\cite{R1}, in the context of quantum mechanics (one-dimensional quantum field theory), non-Hermitian \cPT- symmetric theories~\cite{R2} were shown to allow unitary time evolution.\footnote{A new inner product on the Hilbert space is used which replaces the conventional Dirac inner product.}~$\mathcal P$ is a linear idempotent operator (such as parity) and $\mathcal T$ is an anti-linear idempotent operator (such as time-reversal). In \cPT- symmetric quantum mechanics the energy eigenvalues are real and bounded below. 
~This development has led to the study of \cPT-symmetric quantum field 
theories~\cite{scalarPT,R8a,applic,bjr,ptneutr,R10,R15a,R15,R11,ptsusy}.\footnote{A more general form of non-Hermiticity, known as pseudohermiticity, is discussed in the  Appendix and also can lead to unitary time evolution~\cite{R3,R4} on a Hilbert space with an unconventional inner product.} \end{itemize}
    A major driver for the explosion of interest in quantum mechanical \cPT symmetry has been the massive activity, both theoretical and experimental, in material science and optics~\cite{R4a}.

It is the purpose of this work to discuss the foundations of a (3+1)-dimensional \cPT-symmetric  quantum field theory of fermion and axion fields, from the point of view of its renormalization and dynamical mass generation for both axions and fermions.     
The structure of the article is as follows: in section \ref{sec:model} we set up in detail the path-integral formalism describing the quantum field theory of our model, paying special attention to its \cPT-symmetric nature. In section \ref{sec:yuk}, we present the renormalisation, in (3+1)-dimensions, of our field-theoretic model,  which involves a chiral Yukawa interaction of a fermion field $\psi$ with a pseudoscalar (axion-like) field $\phi$, in the presence of a quartic self-interaction for $\phi$. The fields have bare masses. 
In section \ref{MYukawa}, we discuss the renormalisation group for this massive Yukawa theory, allowing for appropriately defined non-Hermitian fixed points in the space of couplings of the model. We study the behaviour of the various (perturbative) coupling parameters at one loop, derive the beta functions of the renormalisation group (RG), determine the RG fixed points  and study their stability. We also discuss the RG flows of the couplings and masses. In section \ref{sec:NJL}, following the approach of  Nambu and Jona-Lasinio~\cite{R16, R16a} (who considered a non-renormalisable model with quartic fermion interactions, a prototype for dynamical mass generation for fermions), we study the dynamical mass generation in our model, by letting the bare mass terms go to zero. 
We compare the resultant masses with those from the nonperturbative ones obtained  in \cite{R15a,R15} following Schwinger-Dyson (SD) methods (in the absence of the quartic scalar coupling). In section \ref{sec:2loop} we discuss briefly the results of the renormalisation of the model at two loops and demonstrate a renormalisation-group flow also from Hermitian to non-Hermitian couplings. Finally conclusions and outlook are given in section \ref{sec:concl}. A technical discussion on pseudo-Hermiticty, giving background essential for understanding \cPT symmetry and non-Hermiticity, is given in the Appendix.

    \section{Our model, motivation and formulation \label{sec:model}}
       Our axion field theory is a generalisation of the simplest scalar quantum field theory with a non-Hermitian \cPT-symmetric potential  defined by the $D$-dimensional Hamiltonian density:
       \be
       \label{E1}
       H_{PT}=\frac{1}{2} \left( \nabla \phi \right)^{2}  +\frac{1}{2} m^{2}\phi^{2} +g\phi^{2} \left( i\phi \right)^{\delta }.  
       \ee
      $\phi $ is a pseudoscalar field and $ \delta >0$ is real. $H$ is non-Hermitian but \cPT-symmetric because $\phi$ changes sign under $\mathcal P$, $\phi$ remains unchanged under $\mathcal{T}$ and $i$ changes sign under $\mathcal T$. This Hamiltonian density is the field-theoretic analogue of the \cPT-symmetric quantum mechanical Hamiltonian~\cite{R1} 
        \be
        \label{E2}
        H=p^{2}+x^{2}\left( ix\right)^{\delta }  
        \ee
  which launched the field of \cPT symmetry. Dorey {\it et al.}  \cite{R4b, R4c}   demonstrated the surprising feature that for $ \delta>0$ the eigenvalues of $H$ are all discrete, real and positive even though it is not Dirac Hermitian.
  It is necessary to broaden the class covering the Hamiltonian Eq.\eqref{E1} in order to be able to apply the ideas of \cPT symmetry to phenomenologically interesting models.\footnote{There is no proof that, in all cases for which a \cPT symmetry can be defined, the spectrum is purely real. The spectrum will depend on the precise boundary conditions in the problem. If all the energy spectrum is not real, the symmetry is said to be broken. However for a large class of models it has been found that the spectrum is real (and bounded below).} Furthermore it is now realised that \cPT symmetry is part of  a much \emph{broader} class of models  which is denoted as pseudoHermitian~\cite{R3} (see Appendix \ref{sec:appA}).
  
  Recently we have discussed dynamical mass generation~\cite{R15a,R15, R11} for fermions and pseudoscalar  gravitational axion fields in effective field theories containing Yukawa type interactions between the axions and the fermions. These models arise in scenaria for radiative Majorana sterile neutrino masses~\cite{R15}. These Yukawa interactions can be both Hermitian and non-Hermitian but \cPT-symmetric.  Although motivated by the issue of dynamical mass generation, a nonperturbative phenomena, our emphasis in this work is on  understanding  the model within the context of a more fundamental non-Hermitian quantum field theory where the effects of renormalisation need to be considered. A more fundamental model goes beyond the Yukawa interactions in the effective theory to include quartic and cubic couplings in the scalar field. This leads us to consider the generalised Lagrangian $\mathcal{L}$  :
   \begin{equation}
\label{ee1}
\mathcal{L}= \frac{1}{2} \partial_{\mu} \phi \partial^{\mu} \phi -
\frac{M^2}{2} \phi^2 + \bar{\psi} \left( i \slashed{\partial} - m\right)
\psi - i g \bar{\psi} \gamma^5 \psi \phi+ \frac{u}{4!} \phi^{2} \left( i\phi \right)^{\delta } 
=L_{B} +L_{F},
\end{equation}
where 
\be\label{ee1b}
L_{B}=\frac{1}{2} \partial_{\mu} \phi \partial^{\mu} \phi -
\frac{M^2}{2} \phi^2 + \frac{u}{4!} \phi^{2} \left( i\phi \right)^{\delta } ,
\ee
and
\be\label{ee1f}
L_{F}= \bar{\psi} \left( i \slashed{\partial} - m\right)\psi- i g \bar{\psi} \gamma^5 \psi \phi
\psi.
\ee
The scalar model of~Eq.\eq{E1} is contained within this Lagrangian as $L_{B}$.
In \cite{R15}  the analysis was done in the spirit of effective Lagrangians and so the quartic coupling, which emerges from the requirement of renormalisability, was not included.  When we discuss renormalisability we require that $\delta=2\,.$\footnote{Renormalisability for general $\delta$ is discussed in \cite{R7, R8}.} Eq. \eq{ee1} represents the most general renormalisable Lagrangian involving our $\phi$ and $\psi$ fields in four dimensions. For $u<0$ the the quartic interaction is Hermitian; for $u>0$ the quartic coupling is non-Hermitian.\footnote{This remark has to be understood within the context of the deformation implied by $\delta$ and boundary conditions in path integrals.} At the non-pertubative level \cPT symmetric quartic scalar Hamiltonians lead to one-point and, more generally, odd-number point Greens functions.When  $\delta=1$,  $\mathcal{L}$ will have a term $ih\phi^{3} $ where $h$ is a coupling constant; unlike $\phi^{3} $ theory with a real coupling such a theory is~\cPT symmetric and is a sensible theory. 

In order to define a quantum theory, whether we do this through a Schr\"odinger equation (for quantum mechanical theories) or more generally through path integrals for $D$-dimensional quantum field theories we need to \emph{specify boundary conditions}. So if we analytically continue a coupling in some way (the particular sense is specified for the relevant coupling) it is absolutely necessary to show how the boundary conditions are affected. Once these boundary conditions are determined then it will be clear what  Hermiticity or non-Hermiticity means.

The introduction of fermions in a \cPT context needs a discussion and is a comparatively unexplored area within the study of \cPT quantum field theories. For applications of \cPT symmetry to fundamental physics it is important to incorporate fermions~\cite{R15a,R15,R11,R12, R13, R14aa, chern,ptsusy}. In this new area it is not possible currently to match the rigour applied to conventional  Hermitian field theories. However we plan to lay some foundations. The discussion of Feynman rules, $\mathcal P$  and $\mathcal T$ symmetries are connected to the issue of path integrals~\cite{R23} and their boundary conditions. In the context of Lorentz invariant pseudoHermitian field theories, we shall touch on the role $\mathcal{CPT}$ symmetry~\cite{R23a} (where $\mathcal C$ is the charge conjugation operator), and is a fundamental symmetry of Hermitian Lorentz invariant field theories.
The understanding of the role of \cPT~type symmetry in relativistic quantum mechanics and field theory for fermions is less developed than for the bosonic case.  All the above mentioned \cPT-symmetric field-theoretic systems are relativistic, and for which the generation of real masses can, in principle, be understood as a consequence  of the existence of an underlying antilinear symmetry~\cite{most,maninnerPT,antilin}.

\subsection{Bosonic path integrals and boundary conditions \label{sec:Bdry}}
We shall start off in the simplest context: bosonic path integrals with discrete $\mathcal P$ and $\mathcal T$ symmetries.
The  action that will be considered is of the following type
 \be
S\left( \varphi \right)  =\int d^{D}x\left( \frac{1}{2} \  \left( \partial_{\mu } \varphi \right)^{2}  +V\left( \varphi \right)  \right) . 
\ee
 The canonical form of $V\left( \varphi \right) $ used in the study of \cPT~symmetry is
\be
\label{e2}
V\left( \varphi \right)  =\frac{u}{4!} \varphi^{2} \left( i\varphi \right)^{\delta }  
\ee
with $u$ and $\delta$ real. The action of  \cPT~on $V\left( \varphi \right)$ is determined through:
 \begin{align}\label{cpttrn}
  \mathcal{P}&:\quad \varphi \longrightarrow -\varphi \nonumber \\
  \mathcal{T}&: \quad  \varphi \longrightarrow \varphi \nonumber \\
  \mathcal{T}&:\quad i \longrightarrow -i \,.
\end{align}
The potential  $V\left( \varphi \right)$ is \cPT~symmetric for all values of $\delta$.~\cPT is an example of an antilinear symmetry (since $\mathcal{T}$ is antilinear). Pseudo-Hermiticity relies on the presence of an antilinear symmetry. For $\delta=2$  we have the negative quartic potential which is conventionally an unstable potential and energies of states have an imaginary part. The above \cPT~symmetric formulation, involving a complex deformation of the potential, leads to a theory in $D=0$ and $D=1$ with real energies. There are strong grounds to expect this to hold for $D>1$. The purpose of this section is to formulate the analysis in $D=0$ in such a way that the generalisation to $D>0$ is clear (but may have complications such as  renormalisation). The path in $\varphi$ space, because of the deformation parametrised by $\delta$,  is taken (and required) to explore the complex $\varphi$-plane. The presence of \cPT~symmetry results in a left-right symmetry of the deformed path, i.e. a reflection symmetry in the imaginary $\varphi $-axis. This left-right symmetry is responsible for real energy eigenvalues. If, for example,  we have $\mathcal{T}:\  \varphi \longrightarrow -\varphi$ then we do not have \cPT symmetry for general $\delta$, the boundary conditions are different and the left-right symmetry of the deformed paths no longer holds. If the Lagrangian (e.g. for $\delta=2$) formally shows \cPT symmetry for $\mathcal{T}:\  \varphi \longrightarrow -\varphi$ the physical consequences of the different assignments of $\mathcal P$ and $\mathcal T$ are entirely different; one case may give an acceptable physical theory with left-right symmetry and real eigenvalues, while the other case with up-down symmetry would not have real eigenvalues which are bounded below. We will consider below the Euclidean version of the path integral to improve  the convergence of the path integral.

%  \subsubsection{The $D=0$ case}
  
\subsubsection{The quartic potential \label{sec:quartic}}
The partition function for $D=0$ has the form
\be
\label{e3}
Z=\int_{C} d\varphi\,_{} \exp\left(-\left(\frac{1}{2}m^2\varphi^2-\frac{1}{4!}u \varphi^4\right)\right).
\ee
$Z$ represents a zero-dimensional field theory~\cite{R2} and the path integral measure is the measure for contour integration. The study of this toy model (which can formally be investigated as a field theory with Feynman rules)  will help in understanding the role of Stokes wedges~\cite{R20} in path integrals.  For $u>0$ the integral with the contour $-\infty <\varphi <\infty $ does not exist. For $u<0$ the integral with the contour exists in the Stokes wedges $-\frac{\pi }{8} <\arg \varphi <\frac{\pi }{8} $ and $\frac{7\pi }{8} <\arg \varphi <\frac{9\pi }{8} $. Hence the \emph{conventional} Hermitian theory can use the contour $-\infty <\varphi <\infty $ which goes through the centre of both Stokes wedges. It is straightforward to see that there are $4$ possible Stokes wedges each with an opening of $\pi /4$. 
In a \cPT-symmetric context the partition function can exist for a  contour ${C}$ in the complex $\varphi$-plane, chosen to lie in the  Stokes wedges: $-\frac{3\pi }{8} <\arg \varphi <-\frac{\pi }{8}$ and $-\frac{7\pi }{8} <\arg \varphi <-\frac{5\pi }{8} $. These Stokes wedges are left right symmetric and so the \cPT symmetric theory has real eigenvalues which are bounded below.

\subsubsection{The cubic potential \label{sec:cubic}}

An analysis similar to that for the quartic potential can be carried out for the cubic potential~\cite{R20abc} partition function $Z_{3}^{PT} $
\be
\label{E3}
    Z_{3}^{PT} =\int_\mathcal{C} d z\, \exp{-\left(\frac{1}{2}m^2 \varphi^2+\frac{i}{3!}\tilde g \varphi^3\right)}\,.
\ee
where $\tilde g$ is real.~The associated Stokes wedges are $\left\{ -\frac{\pi }{3} <\arg \left( \varphi \right)  <0\right\}$  and $\left\{ -\pi <\arg \left( \varphi \right)  <-\frac{2\pi }{3} \right\}$ and the integral converges along the real $\varphi$ axis. If $\varphi$  is $\mathcal T$-odd, the \cPT conjugate of $\left\{ -\frac{\pi }{3} <\arg \left( \varphi \right)  <0\right\}$ is $\left\{0 <\arg \left( \varphi \right)  <\frac{\pi }{3}\right\}$. The two Stokes wedges are contiguous and so the contour $\mathcal C$ can be deformed off to $\infty$ and the theory would be trivial.

From the above discussions it should be clear that in the presence of both cubic and quartic potentials the Stokes wedges are determined by the quartic potential.
\begin{figure}[!h]
\centering\includegraphics[width=4.5in]{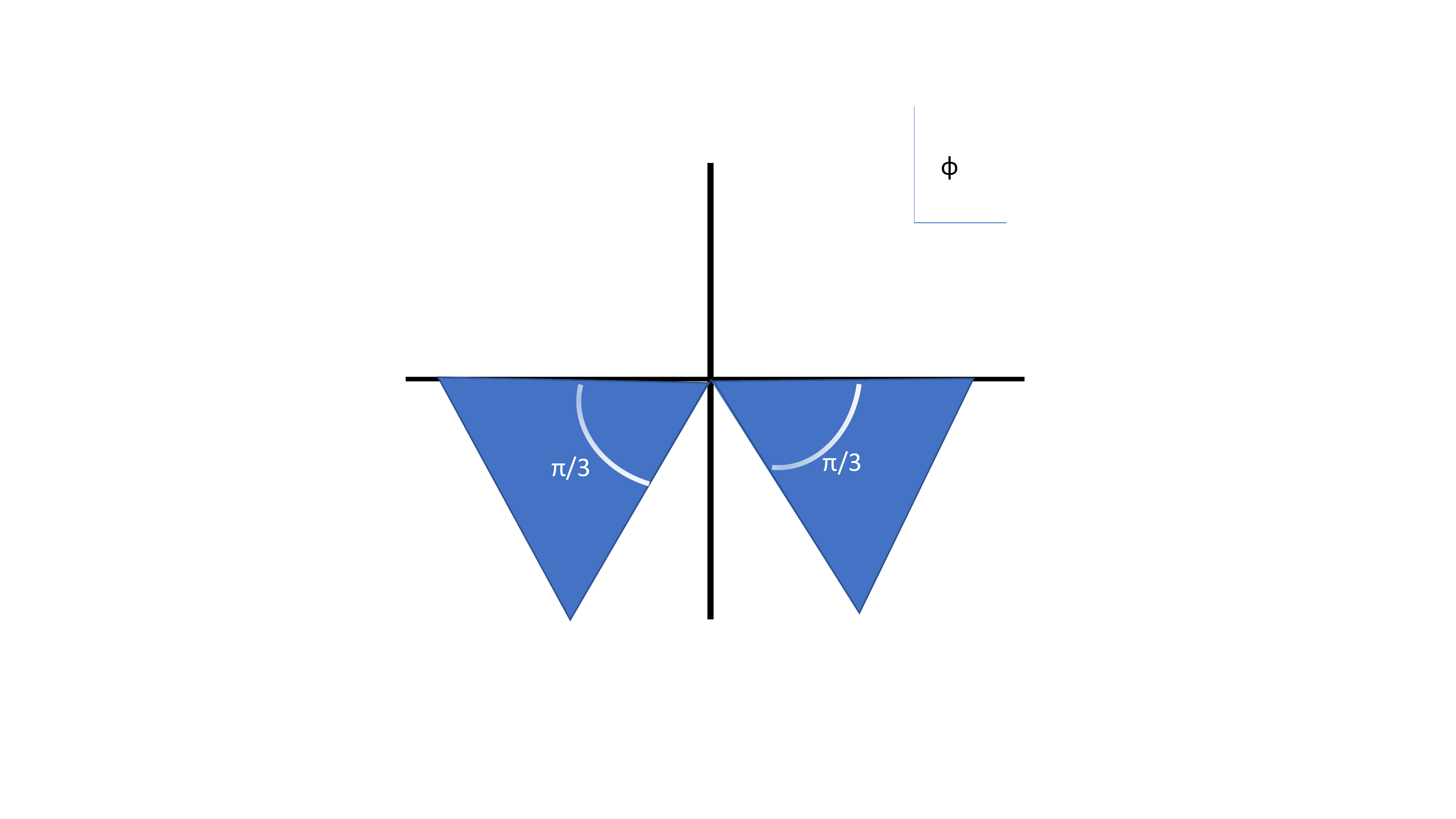}
\caption{ \cPT symmetric Stokes wedges for cubic potential}
\label{fig:diagCubic}
\end{figure}

\begin{figure}[!h]
\centering\includegraphics[width=4.5in]{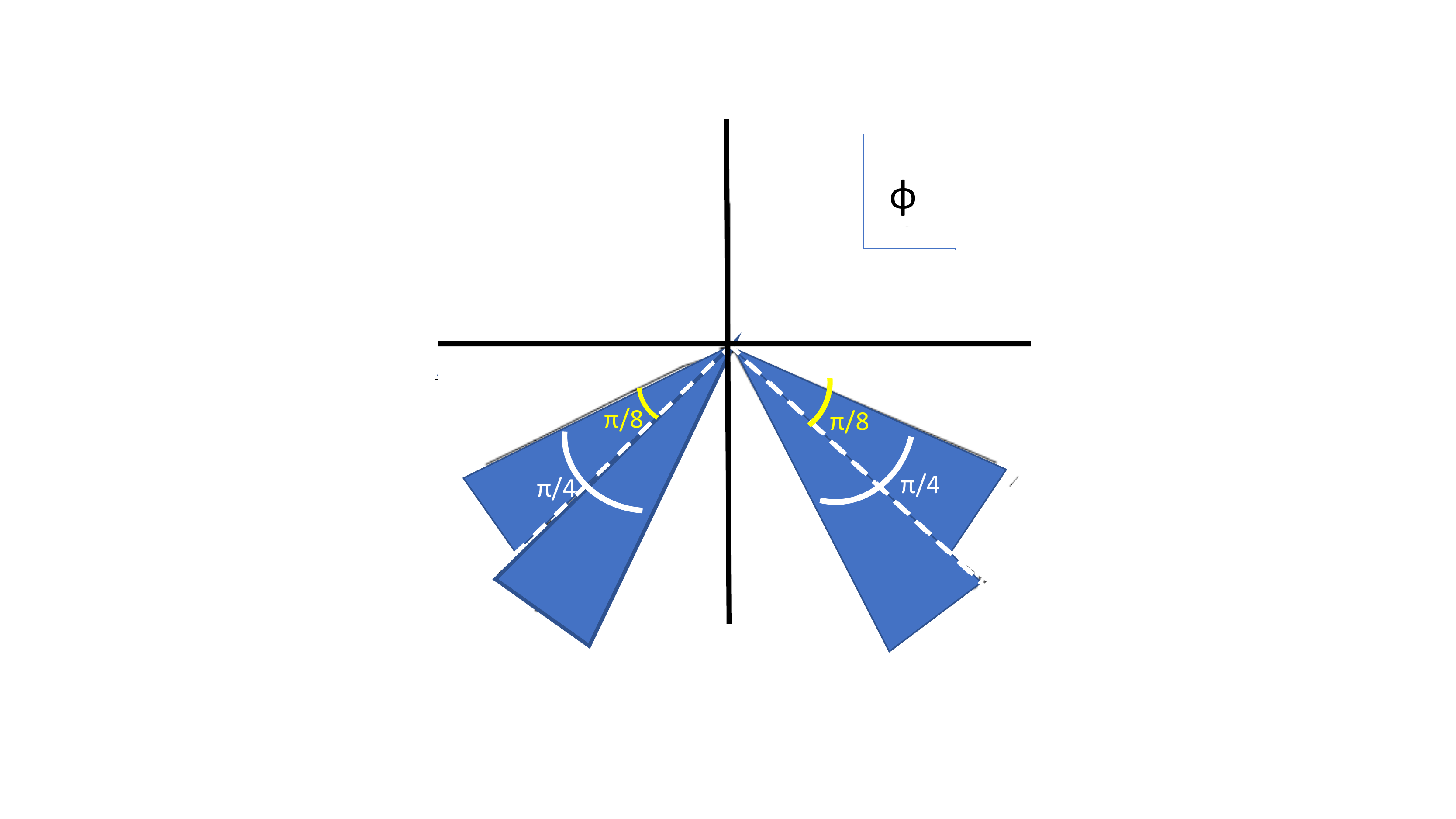}
\caption{ \cPT symmetric Stokes wedges for quartic potential}
\label{fig:diagQuartic}
\end{figure}

\subsubsection{Fermionic path integrals and their role in \cPT symmetry \label{sec:fermionic}
 }

An essential feature of our model is the presence of fermions~\cite{R20ab}. Since our method of analysis is based on path integrals we need to estimate whether the findings on bosonic path integrals are modified by the presence of fermions. The fermionic part of the path integral is in terms of Grassmann numbers which are anticommuting numbers and so Gaussians of Grassmann numbers truncate; at this level there should not be any additional convergence issues in the fermionic theory. To investigate further, since fermions appear quadratically in $L_F$, they can be formally integrated out in the partition function $Z_{eff} $ associated with Eq.\eq{ee1}:
\be
Z_{eff}=\int D\phi \exp \left[ -S_{B}\left( \varphi \right)  \right]  \det \left( \gamma^{\mu } \partial_{\mu } +im+ig\gamma_{5} \varphi \right)  
\ee
where 
\be
\det_{} \left( \gamma^{\mu } \partial_{\mu } +im+ig\gamma_{5} \varphi \right)  =\int D\psi^{\dag } D\psi \exp \left( -\psi^{\dag } \left[  \gamma^{\mu } \partial_{\mu } + im +ig \gamma_5  \varphi \right]  \psi \right).
\ee  
These fermionic determinants have been widely studied using Feynman-diagram representations (see Figs.~\ref{fig:diagVertex} and \ref{fig:diagDeterminant}), and are complicated. 

\begin{figure}[!h]
\centering\includegraphics[width=4.5in]{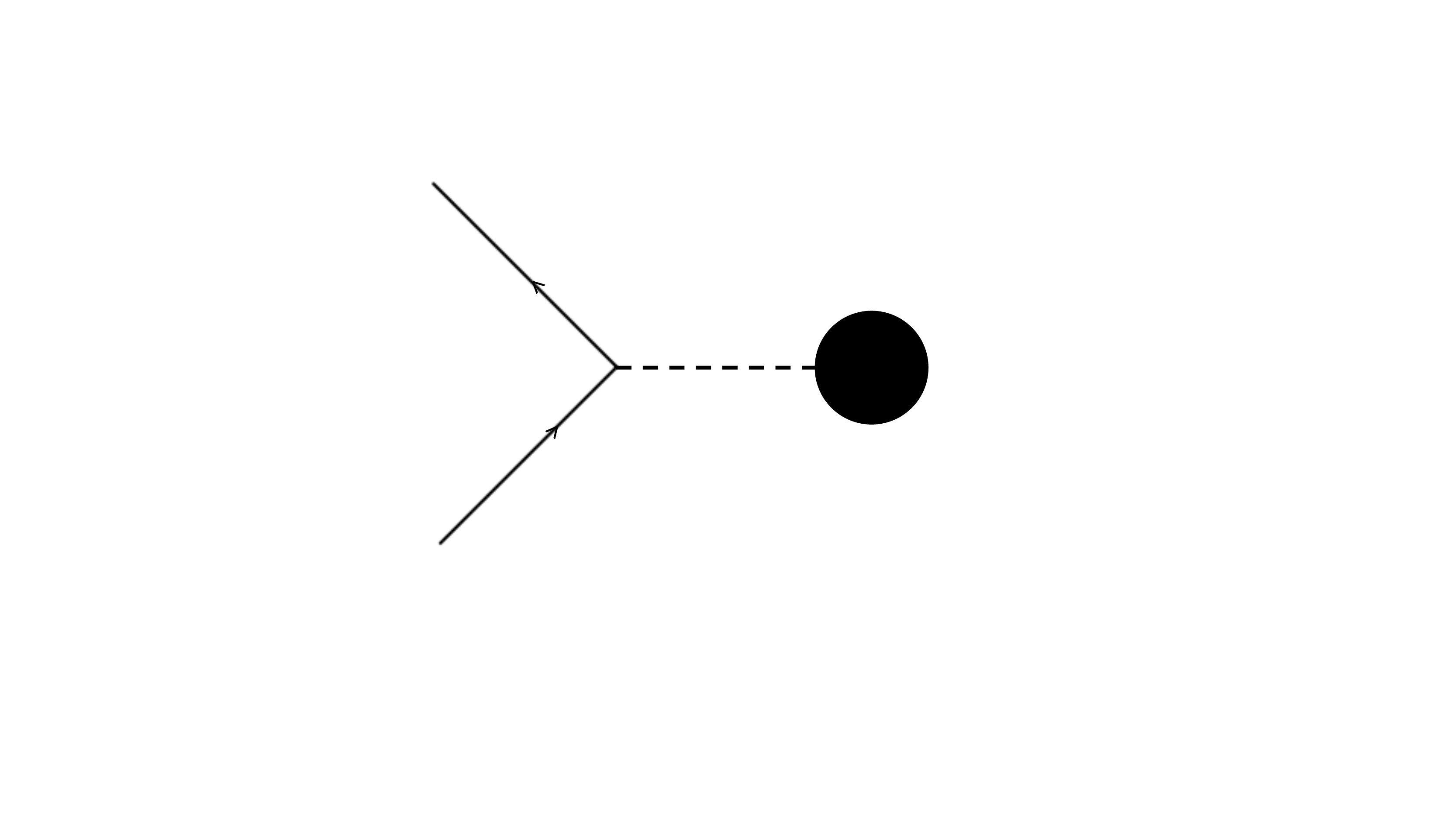}
\vspace{-2.0cm}
\caption{ The master vertex for functional determinant. Continuous lines with arrows denote fermions. The dashed line ending in the dark blob denotes 
an external scalar field source.}
\label{fig:diagVertex}
\end{figure}

\begin{figure}[!h]
\centering\includegraphics[width=4.5in]{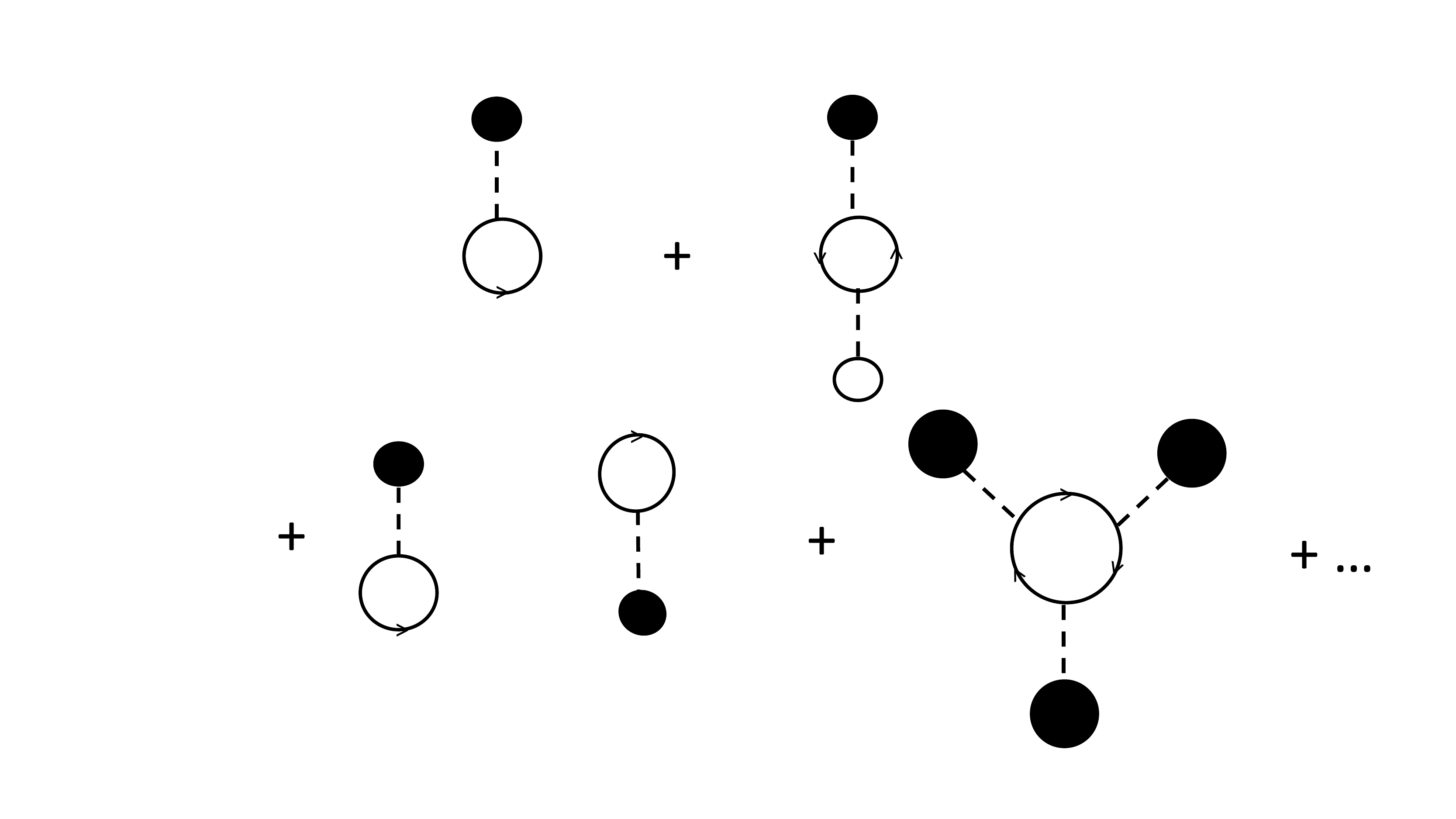}
\caption{Lowest functional vertices for the determinant, including disconnected graphs. The symbols are as in fig.~\ref{fig:diagVertex}.}
\label{fig:diagDeterminant}
\end{figure}

The formal expressions for these determinants are generally non-local; these determinants are approximated using semi-classical methods but even this is non-trivial to do rigorously. On making approximations there is an indication that corrections to the bosonic part of the Lagrangian is of of the form $-u^{2}\varphi^{4} $ and $g^{4} \varphi^{4}$~\cite{R20ab}. Consequently quantum fluctuations may lead to non-Hermitian behaviour even if the starting Lagrangian is Hermitian~\cite{R9}. This issue has relevance within the context of the renormalisation group.

\subsection{The measure of the path integral }\label{sec:m}

The path integral formulation is regularly used to compute correlation (Schwinger) functions in conventional quantum field theories involving scalar, vector and spinor fields. The advantage of this approach is that there is no need to construct a Hamiltonian, the Hilbert space and equation of motion. For these very same reasons  this approach is being advocated by us for \cPT-symmetric field theories and is in our opinion the way forward.
However in the earlier discussion we have been somewhat non-specific about details of the path integral measure except to assume that the properties that we are accustomed to in contour integration and complex analysis continue to serve us well. Since many works on \cPT quantum mechanics rely on Hilbert space methods and modified inner products in the non-Hermitian \cPT-symmetry context, we will give additional supporting arguments for the path integral approach which makes a connection with the Hilbert-space methods used in discussing pseudo-Hermiticity (see Appendix \ref{sec:appA}).

\subsubsection{The calculation of Green's functions}

Since the Dirac inner product of quantum mechanics (the $L^{2}$ norm), when applied to $\mathcal{PT}$-symmetric Hamiltonians leads to a nonunitary quantum-mechanical theory, in the canonical operator approach unitarity is restored through the introduction of a modified inner product~\cite{R3}.\footnote{The formulations in terms of either pseudo-Hermiticity (see appendix \ref{sec:appA}) or an invertible metric operator or a $\mathcal{C}$-operator ~\cite{R20abc} are entirely equivalent. }
Unlike the inner product in Hermitian quantum mechanics, this modified inner product is \emph{not} uniquely determined and is dependent  on the Hamiltonian.
Through examples we shall compare the calculation of a two-point function in the path integral and canonical approaches . Such  examples provide evidence for the conjecture that for the calculation of Green's functions (within both a Minkowski and an Euclidean framework), the determination of the  $\mathcal{C}$ -operator (or equivalently the Hilbert space metric) is not necessary (at least for a class of models where the non-Hermiticity lies in the interaction part of the Hamiltonian). This evidence stimulated a formal justifications of these findings~\cite{R6abc}.

The evidence is based on both an exactly soluble $\mathcal{PT}$ symmetric quantum mechanical model (the Swanson model) and also on an a perturbative treatment of the imaginary cubic potential~\cite{R20abc}.  
We shall then outline an  argument which justifies, in a general context,  conclusions deduced from these two models.

\vspace{.1cm}

The $\mathcal{C}$-operator~\cite{R22} can be written as 
\begin{align}\label{Cop}
\mathcal{C}=\exp\left(\mathcal{Q}\right)\mathcal{P}\,,
\end{align} 
where $\mathcal{Q}$ is Hermitian and $\mathcal{P}$ is the linear parity operator~\cite{R2}. Moreover $\eta=\exp\left(\mathcal{-Q}\right)$ where $\eta$ is the metric  in the pseudo-Hermitian formulation of $\mathcal{PT}$ symmetry. The conventional adjoint $H^{\dagger}$ of a Hamiltonian $H$ satisfies
\be
\label{ea1}
H^{\dagger}=\exp\left(\mathcal{-Q}\right)H\exp\left(\mathcal{Q}\right)
\ee 
and this leads to an associated Hermitian Hamiltonian $h$ which is defined as 
\be
\label{ea2}
h=\exp\left(\mathcal{-\frac{Q}{\mathit{2}}}\right)H\exp\left(\mathcal{\frac{Q}{\mathit{2}}}\right).
\ee

\subsection {\nt{The Swanson model \label{sec:B1}}}
The classical Swanson Hamiltonian (with $a>0, b>0$ and $c$ pure imaginary)~\cite{R5,R6abc} $H_{S}$ is
\be
\label{ea3}
H_{S}=ax^{2}+bp^{2}+2cxp
\ee
from which we deduce (upto a total derivative) a classical Lagrangian $L_{S}=\frac{\dot{x}^{2}}{4b}-\tilde{a}x^{2}$ where $\tilde{a}=a-\frac{c^{2}}{b}$. $L_{S}$ is a scaled Lagrangian for a Hermitian harmonic oscillator and leads to a Hermitian Hamiltonian
\be
\label{ea4}
h_{S}\left(x,P\right)=bP^{2}+\tilde{a}x^{2}
\ee 
with
$P=p+\left(c/b\right)x$. Using conventional techniques such as path integrals for Hermitian Hamiltonians, the time ordered $n$-point Greens functions $G_{n}\left(t_{1},t_{2},\ldots,t_{n}\right)$ can be calculated. 

From the theory of pseudo-Hermitian Hamiltonians there is a  similarity transformation (determined by a $Q$ function) which relates the two.   The ground states of the two Hamiltonians are also related by this similarity transformation:

\be
\label{ea5}
\left|\Omega_{h_{S}}\right\rangle =\exp\left(\mathcal{-\frac{Q}{\mathit{2}}}\right)\left|\Omega_{H_{S}}\right\rangle .
\ee
Since $G_{2}\left(t,t\right)=\left\langle \Omega_{h_{S}}\right|x^{2}\left|\Omega_{h_{S}}\right\rangle $, we can rewrite it as
\be
\label{ea6}
G_{2}\left(t,t\right)=\left\langle \Omega_{H_{S}}\right|\exp\left(-\frac{\mathcal{Q}}{2}\right)x^{2}\exp\left(-\frac{\mathcal{Q}}{2}\right)\left|\Omega_{H_{S}}\right\rangle  .
\ee
Moreover, in this case, it is possible to find a $\mathcal{Q}$ which depends just on $x$, and so
\be
\label{ea7}
G_{2}\left(t,t\right)=\left\langle \Omega_{H_{S}}\right|\exp\left(-\mathcal{Q}\right)x^{2}\left|\Omega_{H_{S}}\right\rangle,
\ee 
the form expected in a non-Hermitian framework. Hence in this simple case a path integral computation and one using a $\mathcal{PT}
$-symmetric framework with a $\mathcal{PT}$ inner product give the same result.

\subsection{\nt{The imaginary cubic potential}\label{sec:imaginary}}
An example of a model which is not exactly soluble but can be                                                                                                                                                                                                                                                          solved by using perturbation theory is provided by the Hamiltonian
\be
\label{ea8}
H_{C}=\frac{1}{2}p^{2}+\frac{1}{2}x^{2}+i\tilde gx^{3}
\ee
with $\tilde g$ real.
 
The associated classical Lagrangian is
\be
\label{ea9}
L_{C}=\frac{1}{2}\left(\dot{x}^{2}-x^{2}\right)-i\tilde gx^{3}.
\ee
From our earlier discussion of Stoke's wedges, we know that the edge of the Stoke's wedges coincides with the real $x$ axis; so conventional Feynman rules can be used valid and lead to~\cite{R6a}
 \be
\label{ea10}
G_{1}=-\frac{3}{2}i\tilde g+\frac{33}{2}i\tilde g^{3}+O(\tilde g^5).
\ee
This result coincides with the quantum mechanical calculation using $H_{C}$ and the  $\mathcal{PT}$-symmetric inner product, i.e. $\left\langle 0\right|\exp\left(-\mathcal{Q}\right)x\left|0\right\rangle $ for a suitable operator $\mathcal{Q}$~\cite{R6aa}.~Hence we have further evidence supporting the conjecture which prompted the investigation of Jones and Rivers~\cite{R5} to be discussed next.

\subsection{\nt{General argument} \label{sec:general}}

A quantum field theory is characterised by its Green's functions. The Schwinger-Dyson equations (SDE)~\cite{R23}, which are c-number equations, determine the Green's functions of a field theory.The SDE can be derived from the partition function $Z[j]$ where $j(x)$ denotes a source field. For definiteness we will consider a pseudoscalar field $\phi (x)$  (in a spacetime dimension $D$) with an action $S\left[\phi\right]=\int L\left(\phi\left(x\right),\partial_{\mu}\phi\left(x\right)\right)d^{D}x$ where $L\left(\phi\left(x\right),\partial_{\mu}\phi\left(x\right)\right)$ is the Lagrangian density. We can take for definiteness
\be
\label{ea11}
L\left(\phi\left(x\right),\partial_{\mu}\phi\left(x\right)\right)=\frac{1}{2}\partial_{\mu}\phi\partial^{\mu}\phi-\frac{1}{2}m^{2}\phi^{2}-U\left(\phi\right)
\ee
where $U\left(\phi\right)$ is in general a non-Hermitian Hamiltonian.
$Z[j]$ can be represented in two ways: one in terms of a path integral denoted by $Z_{1}[j]$  and the other in terms of a time-ordered product of operators  denoted by $Z_{2}[j]$. We shall use the relation between the two expressions to argue that the path integral approach does not require, as far as the computation of Green's functions is oncerned, explicit knowledge of the non-Hermitian metric.
The expressions for $Z_{1}[j]$ and $Z_{2}[j]$ are
\be
\label{ea12}
Z\left[j\right]=Z_{1}[j]=\int D\phi\exp\left(-S\left[\phi\right]+\int j\left(x\right)\phi\left(x\right)\right)
\ee
and
\be
\label{ea13}
Z\left[j\right]=Z_{2}[j]=\left\langle \varOmega\right|\eta T\left(\exp\left[i\int dxj\left(x\right)\phi\left(x\right)\right]\right)\left|\Omega\right\rangle,
\ee 
where $\left|\Omega\right\rangle$ denotes the vacuum state. The metric operator $\eta$ is time-independent.

From \eq{ea12} we obtain the SDE on requiring that  
\be
\label{ea14}
\int D\phi\frac{\delta}{\delta\phi\left(x\right)}\exp\left[-S\left[\phi\right]+\int dy\phi\left(y\right)j\left(y\right)\right]=0. 
\ee
The Green's functions are obtained from 
\be
\label{ea15}
G_{n}\left(x_{1},x_{2},\ldots,x_{n}\right)=\frac{1}{Z[j]}\left(-i\frac{\delta}{\delta j\left(x_{1}\right)}\right)\left(-i\frac{\delta}{\delta j\left(x_{2}\right)}\right)\ldots\left(-i\frac{\delta}{\delta j\left(x_{n}\right)}\right)Z\left(j\right)\vert_{j=0}.
\ee
The path integral measure is formally encoded in $D\phi$, but, in the derivation of the SDEs, it is not necessary to specify this measure precisely.  The main assumption is that the path integral exists. From \eq{ea14} we deduce that 
\be
\label{ea16}
\left(-\frac{\delta S}{\delta\phi\left(x\right)}\vert_{\phi\left(x'\right)=\frac{\delta}{\delta j\left(x'\right)}}+j\left(x\right)\right)Z\left(j\right)=0.
\ee
Alternatively we can derive the SDEs using $Z_{2}[j]$. The derivation  starts from the Heisenberg equations of motion. ${\mathcal{H}}$  is given by
\be
\label{ea17}
{\mathcal{H}}=\int dx\left[\frac{1}{2}\pi^{2}+\frac{1}{2}\left(\nabla\phi\right)+\frac{1}{2}m^{2}\phi^{2}+U\left(\phi\right)\right]
\ee
with $\pi=\partial_{0}\phi$ and we assume $\left[{\mathcal{H}},\eta\right]=0$. The Heisenberg equations of motion (in natural units and using notation which does not distinguish between classical and operator fields) can be shown to be:
 \be
 \label{ea18}
 \left(\partial^{2}+m^{2}\right)\phi\left(x\right)+U'\left(\phi\left(x\right)\right)=0
 \ee
 where 
 \be
 \label{ea19a}
 \partial_{0}\phi=i\left[{\mathcal{H}},\phi\right]
 \ee 
 
 and 
 
 \be
  \label{ea19b}
 \partial_{0}\pi=i\left[{\mathcal{H}},\pi\right]. 
 \ee 
 
 The assumption that these Heisenberg equations (\eq{ea19a},\eq{ea19b}) are valid for a pseudo-Hermitian Hamiltonians $\mathcal{H}$ will now be justified.
 For ${\mathcal{H}}$ to be pseudo-Hermitian with respect to an inner product $\eta$ (discussed earlier) a necessary condition is that
  \be
  \label{ea20}
 \eta\exp\left(i\mathcal{H}\right)=\exp\left(i\mathcal{H}^{\dagger}\right)\eta.
  \ee
 In Hilbert space a ket  $\left|\psi_{S}\right\rangle $ in the Schr$\ddot{\rm o}$dinger picture is related to a ket $\left|\phi_{H}\right\rangle $ in the Heisenberg picture by
 \be
  \label{ea21a}
\left|\phi_{H}\right\rangle =\exp\left(i\mathcal{H}t\right)\left|\psi_{S}\right\rangle. 
\ee 
Similarly for the corresponding bras
\be
\label{ea21b}
\left\langle \phi_{H}\right|=\left\langle \psi_{S}\right|\exp\left(-i\mathcal{H}^{\dagger}t\right).
\ee
For an operator $\mathcal{O}$ in the Schr$\ddot{\rm o}$dinger picture
\begin{eqnarray}
\label{ea22}
\left\langle \psi_{S}\right|\eta\mathcal{O}\left|\psi_{S}\right\rangle  & = & \left\langle \phi_{H}\right|\exp\left(i\mathcal{H}^{\dagger}t\right)\eta\,\mathcal{O}\exp\left(-i\mathcal{H\mathit{t}}\right)\left|\phi_{H}\right\rangle  \nonumber \\
 & = & \left\langle \phi_{H}\right|\eta\exp\left(i\mathcal{H}t\right)\mathcal{O}\exp\left(-i\mathcal{H\mathit{t}}\right)\left|\phi_{H}\right\rangle \nonumber \\
  & = & \left\langle \phi_{H}\right|\eta\mathcal{O_{H}}\left|\phi_{H}\right\rangle 
\end{eqnarray}
where 
\be
\label{ea23}
{\mathcal{O}}_{H}=\exp\left(i\mathcal{H}t\right)\mathcal{O}\exp\left(-i\mathcal{H\mathit{t}}\right).
\ee
Hence, even in the case of pseudo-Hermitian Hamiltonians, the Heisenberg picture operator ${\mathcal{O}}_{H}$ obeys the standard form of the Heisenberg equations of motion.

 In the canonical formulation $Z_{2}[j]$, we shall adapt the Symanzik construction for SDE in the presence of a source. The metric appears explicitly in this formulation. In this construction the expectation values of fields are determined by the known classical equations of motion and the equal time commutation relations. If we find that the SDE are the same in the path integral and canonical approaches, then we can deduce that the path integral formulation (without an explicit implementation of $\eta$) leads to a correct calculation of Green's functions in the case of pseudo-Hermitian Hamiltonians. For a given $x^{\mu}=\left(x_{0},\overrightarrow{x}\right)$ (and with all fields below operator-valued)~\cite{R23}
\be
\label{ea24}
Z_{2}[j]=\left\langle \varOmega\right|\eta \left(\mathcal{E}\left(\infty,x_{0}\right)\mathcal{E}\left(x_{0},-\infty\right)\right)\left|\Omega\right\rangle 
\ee
 and formally 
 \be
 \label{ea25}
 \left(-i\frac{\delta}{\delta j\left(x\right)}\right)^{p}Z_{2}[j]=\left\langle \varOmega\right|\eta \mathcal{E}\left(\infty,x_{0}\right)\mathcal{\mathcal{\phi}\mathit{(x)}^{\mathit{p}}E}\left(x_{0},-\infty\right)\left|\Omega\right\rangle .
 \ee
 where
 \begin{equation}
\label{ea25b }
\mathcal{E}\left( x^{\prime }_{0},x_{0}\right)  =T\left[ \exp \left( i\int^{x^{\prime }_{0}}_{x_{0}} dy_{0}\int d\vec{y} j\left( y_{0},\vec{y} \right)  \phi \left( y_{0},\vec{y} \right)  \right)  \right].  
\end{equation} 

 Hence 
  \be
  \label{ea26}
 0=\left\langle \varOmega\right| \eta {\mathcal{E}}\left(\infty,x_{0}\right)\mathcal{\left(-\frac{\delta S}{\delta\mathcal{\phi}\mathit{(x)}}\right)E}\left(x_{0},-\infty\right)\left|\Omega\right\rangle\,, 
 \ee
 and
 \begin{eqnarray}
 \label{ea27}
 \quad \quad 0&=&\left[\left(\partial^{\mathit{2}}+\mathit{m}^{\mathit{2}}\right)\left(-i\frac{\delta}{\delta j\left(x\right)}\right)+U'\left(-i\frac{\delta}{\delta j\left(x\right)}\right)\right]Z_{2}[j] \\
&&+\left\langle \varOmega\right|\eta\mathcal{E}\left(\infty,x_{0}\right)\mathcal{\mathit{\mathit{\partial}}_{\mathit{0}}^{\mathit{2}}\phi\mathit{(x)}E}\left(x_{0},-\infty\right)\left|\Omega\right\rangle -\mathit{\mathit{\partial}}_{\mathit{0}}^{\mathit{2}}\left\langle \varOmega\right|\eta\mathcal{E}\left(\infty,x_{0}\right){\phi\mathit{(x)}E}\left(x_{0},-\infty\right)\left|\Omega\right\rangle \nonumber
 \end{eqnarray}
 The last two terms in \eq{ea27} can be simplified further. We first note that 
 \be \nonumber
\partial_{0}\left\langle \varOmega\right|\eta\left(\mathcal{E}\left(\infty,x_{0}\right)\mathcal{\phi\mathit{(x_{0},\overrightarrow{x})}E}\left(x_{0},-\infty\right)\right)\left|\Omega\right\rangle =\left\langle \varOmega\right|\eta\left(\mathcal{E}\left(\infty,x_{0}\right)\mathcal{\pi\mathit{(x_{0},\overrightarrow{x})}E}\left(x_{0},-\infty\right)\right)\left|\Omega\right\rangle  
 \ee
 since $\phi\mathit{(x_{0},\overrightarrow{x})}$ commutes with itself at equal times. Differentiating again with respect to $x_{0}$ we obtain
 \be
 \label{ea28}
 \partial_{0}^{2}\left\langle \varOmega\right|\eta\left(\mathcal{E}\left(\infty,x_{0}\right)\mathcal{\phi\mathit{(x_{0},\overrightarrow{x})}E}\left(x_{0},-\infty\right)\right)\left|\Omega\right\rangle =\left\langle \varOmega\right|\eta\left(\mathcal{E}\left(\infty,x_{0}\right)\mathcal{\mathit{\partial}_{\mathit{0}}^{\mathit{2}}\phi\mathit{(x_{0},\overrightarrow{x})}E}\left(x_{0},-\infty\right)\right)\left|\Omega\right\rangle +j\left(x\right)
 \ee
 Using \eq{ea28} in \eq{ea27} we finally find \eq{ea16} again but using the inner product $\eta$ this time.
 This demonstration has been confined to scalar theories. In our earlier discussion the fermions did not bring any qualitatively different issues into defining the boundary conditions of the path integral; consequently this derivation should formally still be valid.  

\subsection{\nt{Renormalisation and the path integral measure}\label{sec:renormalisation} }

Recent works~\cite{R14aa, R13, chern} on local \cPT-fermionic quantum field theory have \emph{not} addressed the essential issue of renormalisation~\cite{R9} which arise due to quantum fluctuations. Now that, in the context of \cPT field theories, we have presented a full discussion of the quantisation procedure through path integrals , we will give a detailed analysis of renormalisation and the renormalisation group (RG), within our simple model theory studied as an effective theory in the context of axion physics~\cite{R11, R15} ( involving a Yukawa coupling of a Dirac fermion field $\psi$ to $\phi$, a real pseudoscalar field).
The Yukawa coupling can be real or imaginary. 

Given the possibility that some non-Hermitian field theories may be a basis for fundamental theories, we aim to study the model as a quantum field theory.~The renormalisation group, leads to coupling constants running with energy scale. The connection between Hermitian and non-Hermitian couplings through these flows will be examined.~In addition, \ntt{on noting that the existence of an underlying antilinear (\cPT) (~\cite{R1,R11, R15}) symmetry~\cite{most,maninnerPT,antilin}  in the models, allows for real energy eigenvalues,} we shall examine, using the renormalisability of the model, dynamical mass generation for the fermion and pseudoscalar fields . In the case of small couplings, our  analysis yields non-perturbative results for the generated \ntt{(real)} masses, which  agree with a (one-loop) Schwinger-Dyson (SD) analysis~\cite{R11,R15}~for the model without axion self-interaction. 

\subsection{ Issues on  $\mathbb{C}\mathcal{PT}$\, and $\mathcal{CPT}$ Invariance}

$\mathbb{C}\mathcal{PT} (\equiv \Theta$) invariance~\cite{R23a}, where $\mathbb{C}$ is the conventional charge conjugation operator of Dirac~\cite{BD,schwartz}, should not be confused with the $\mathcal C$ operator \eqref{Cop} discussed in section \ref{sec:m}. This invariance 
has been proved for Hermitian Lorentz invariant theories and is sometimes referred to as the (Hermitian) $\mathbb{C}\mathcal{PT}$ theorem . $\Theta$ is an important symmetry, which relates matter and antimatter: for example the equality of masses and lifetimes, and opposite charges, for particles and antiparticles. It is important to discuss the fate of this symmetry in local relativistic non-Hermitian quantum field theories. This is still an open issue. For the conventional $\mathbb{C}\mathcal{PT} $ theorem, one expects the violation of this symmetry in the presence of general non-Hermitian couplings. Purely phenomenological considerations were adopted in an early study \cite{okun}, to discuss potential experimental searches for general non-Hermitian $\Theta$-violating quantum field theories. 
In what follows we shall concentrate only on our type of Lagrangian \eqref{ee1}. 
For the case of a  purely imaginary  $g$ coupling and positive $u$ the theory is pseudo-Hermitian (see the discussion of pseudo-Hermiticity in the Appendix and boundary conditions on a Feynman path integral in section \ref{sec:Bdry}).

The  Yukawa interaction, if non-Hermitian with the Dirac inner product, turns out to be $\mathbb{C}\mathcal{PT} $ odd~\cite{BD,schwartz}. This can lead, in principle, to observable consequences~\cite{okun}.
 In this work and in \cite{R15,R11}) the antiparticle state can be defined perturbatively in the Yukawa coupling.
However, from a foundational view point it would be interesting to see whether, on restricting to pseudo-Hermitian theories,  one can define in principle a \emph{new}  set of $\mathbb{C}$, $\mathcal{P}$ and $\mathcal{T}$ operators such that the resultant \emph{new} $\Theta$ operator  is an antilinear symmetry of the non-Hermitian theory. 
In a general \cPT-symmetric (or  pseudo-Hermitian) case the Hamiltonian $H$ is not Hermitian and so the conventional antiunitary $\mathbb{C}\mathcal{PT}$ operator  does not map a particle state into an antiparticle state. As noted in section \ref{sec:m} (see also Appendix A),
corresponding to a pseudo-Hermitian $H$ there is a Hermitian $h$ related by a Hermitian Hilbert-space automorphism $\tilde\eta$(=$\eta^{1/2}$ where $\eta$ is defined in Eq. \eq{ea20}) :
\begin{equation}
\label{cpt6}
h=\tilde\eta H\tilde\eta^{-1} .
\end{equation}
The new inner product $\langle \langle .|.\rangle \rangle$ for the non-Hermitian operators  is given by $\langle .|\tilde\eta^{{}^{2}} .\rangle $. 

This statement is strictly valid for finite dimensional quantum  mechanical pseudo-Hermitian systems. Were we to assume the validity of the mapping \eqref{cpt6}, though,  in quantum field theoretic systems, it would be possible in principle (using a similarity transformation $\tilde \eta$ cf. Eq. \eq{simil}) to construct a $\mathbb{C}\mathcal{PT}$ operator that can define the antiparticle state non-perurbatively in pseudo-Hermitian field theories.\footnote{Lattice field theories on a finite lattice are finite dimensional and are a viable regularisation of continuum field theories.  } However the existence of well-defined similarity transformations which lead to a useful $\mathbb{C}\mathcal{PT}$ operator  needs further investigation.

There is another perspective on $\mathbb{C}\mathcal{PT}$ invariance of pseudo-Hermitian field theories, which uses complex Lorentz transformations, see \cite{antilin}, to claim that the conventional $\mathbb{C}\mathcal{PT}$ operator ($\widehat{\theta_{{\mathbb C}\mathcal{PT}}}$) is the correct $\mathbb{C}\mathcal{PT}$ operator  for pseudo-Hermitian Hamiltonians. We believe this assertion to be unproven.  In that work, it is asserted that there are two conditions under which a $\mathbb{C}\mathcal{PT}$ invariance theorem for non-Hermitian systems would be valid. The first is the existence of an antilinear symmetry, which replaces Hermiticity, and ensures the 
time-independence of the appropriate inner products that enter the non-Hermitian theory. The second condition is the extension of the requirement of Lorentz invariance, to encompass invariance under complex Lorentz transformations. This approach is not applicable to our case. In the work of \cite{antilin}, the field operator  $i \overline \psi \, \gamma^5 \, \psi$ is $\mathbb{C}\mathcal{PT}$  picks up a +1 phase (under appropriate  normalization of the phases in the definition of $\mathcal{P}$, $\mathcal{T}$} and  $\mathbb{C}$) under the application of the pertinent transformation. For  conventional  \cite{antilin} $\mathbb{C}\mathcal{PT}$ invariance to hold,  the 
Yukawa  interaction with a pseudoscalar $\phi$ would require a $\mathcal T$-odd transformation of $\phi$ (see \ref{sec:Bdry} ) and a {\it real} coupling. However our term with purely imaginary Yukawa coupling,  is $\mathbb{C}\mathcal{PT}$ odd under our assumed transformations \eqref{cpttrn}; so  the considerations of \cite{antilin} do \emph{not} apply. On the other hand, the non-Hermitian self-interactions of axions in our model satisfy the criteria of \cite{antilin} for $\mathbb{C}\mathcal{PT}$ invariance.

The first criterion of \cite{antilin} , the existence of an antilinear symmetry, such as \cPT, is guaranteed in our case as well, thus leading to either reality of the energy spectrum, or at least appearance of the energy eigenvalues  in complex conjugate pairs.
In what follows we shall examine dynamical mass generation with real eigenvalues in our Yukawa-system with axion self-interactions \eqref{ee1} (\eqref{ee1b},\ \eqref{ee1f}) and demonstrate (in section \ref{sec:NJL}) that this is possible in the model with {\it Hermitian} Yukawa interactions and {\it non-Hermitian} $\mathbb{C}\mathcal{PT}$ even axion self-interaction couplings, under some circumstances, which we shall specify (see also \cite{R11,R15}). 
%We stress once more that the non-Hermitian axion self-interactions are $\mathbb{C}\mathcal{PT}$ even. 
In the non-Hermitian perturbative Yukawa-interaction case, however, as we shall see, dynamical  mass generation, when applied naively, i.e. via the replacement of the real Yukawa couplings by the purely imaginary ones,  
leads to unacceptably large masses (above the UV cutoff), which are thus not self-consistent. 
 It should be noted that in \cite{R11,R15}, for a model with an attractive $\mathbb{C}\mathcal{PT}$-even four-fermion interaction and an anti-Hermitian Yukawa interaction, one can obtain dynamical masses for fermions and pseudoscalars, of approximately equal magnitude 
proportional to $|g| \, \Lambda$ where $\Lambda$ is the ultraviolet cutoff, and $g$ is the Yukawa coupling (as in  \eqref{ee1}). We shall now give a simple argument as to show how this might be understood using the conventional  $\widehat{\theta_{{\mathbb C}\mathcal{PT}}}$.

 In quantum mechanics we have 
 \begin{align}\label{cptstates}
\Big(\widehat{\theta_{{\mathbb C}\mathcal{PT}}}\, \widehat H^\ast  -  \widehat H \, \widehat{\theta_{{\mathbb C}\mathcal{PT}}}\Big) \, |E\rangle = \delta \, \overline{|E\rangle} \ne 0\,,
\end{align}
where $\widehat H$ is a non-Hermitian Hamiltonian operator with an energy eigenstate $|E\rangle$, and  $\ast$ denotes standard complex conjugation and $\overline{|E\rangle}$ denotes the antiparticle energy eigenstate, defined by  $\widehat{\theta_{{\mathbb C}\mathcal{PT}}}\, |E\rangle = \overline{|E\rangle }$.

% $\widehat{\theta_{{\mathbb C}\mathcal{PT}}}$ 
% The question in this $\mathbb{C}\mathcal{PT}$ violating case, 
%s what is the dynamical mass of the appropriate antiparticle states. From a quantum theory viewpoint the difference can be calculated from the 
%action of the (non-vanishing) generalised commutator of the operator corresponding to the appropriate antiparticle-state generator, $\widehat{\theta_{{\mathbb C}\mathcal{PT}}}$ with the (non-hermitian) Hamiltonian operator $\widehat H$ of the system on an energy eigenstate, $|E\rangle$. Let $\overline{|E\rangle}$ denotes the antiparticle energy eigenstate, defined by  $\widehat{\theta_{{\mathbb C}\mathcal{PT}}}\, |E\rangle = \overline{|E\rangle }$.

%where $\dagger$ denotes the standard hermitian conjugate, and $\overline{|E\rangle}$ denotes the antiparticle energy eigenstate, obtained from  $\widehat{\theta_{{\mathbb C}\mathcal{PT}}}\, |E\rangle = \overline{|E\rangle }$. 
%In our case \eqref{ee1}, the only source of non  hermiticity in the sense of $H^\dagger \ne H$ is the purely imaginary Yukawa interaction
%$ i\,  g\,\overline \psi (x) \gamma^5 \, \psi (x)$, with $g=i\breve g$, $\breve g \in \mathbb R$ a real quantity (the pseudoscalar self-interactions are real).
%The existence of an antilinear \cPT symmetry in our system guarantees that the energy eigenvalues $H\, |E\rangle = E\, |E\rangle$ of the non-hermitian $H^\dagger \ne H$ Hamiltonian operator would be either real or come into complex conjugate pairs. Thus from \eqref{cptstates} one would obtain
Hence
\begin{align}\label{cptstates2}
\widehat H \, \overline{|E\rangle } = (E^\star - \delta)\, \overline{|E\rangle }\,,
\end{align}
%where $\star$ denotes complex conjugate.
 In our case, $E=E_1 + i \mu$, with $E_1, \mu \in \mathbb R$. Here, $E_1$ corresponds to the real dynamical masses from  the earlier SD analysis~\cite{R11,R15}, whilst $i\mu \propto \langle \overline \psi \, \gamma^5 \, \psi \rangle$, $\mu \in \mathbb R$ would represent the 
purely imaginary chiral condensate, corresponding to  an antiHermitian chiral mass for the fermions. 

In \cite{R15a} we have argued that it is {\it not} possible, for {\it energetic} reasons, to generate dynamically, using SD analysis, a non-zero non-Hermitian condensate $\langle \overline \psi \, \gamma^5 \, \psi \rangle$, {\it i.e.}, dynamically one should have  $\mu =0$.
We interpret this result as implying that, in the massive phase of the system, the mass eigenvalue of the non-Hermitian operator $\overline \psi \, \gamma^5 \, \psi$ on an energy eigenstate $|E\rangle>$ would vanish 
\begin{align}\label{mueigen}
\overline \psi \, \gamma^5 \, \psi \, |E\rangle =0~.
\end{align}

%If we use the conventional operator $\widehat{\theta_{{\mathbb C}\mathcal{PT}}}$, with ${\mathbb C}$ the 
%conventional charge-conjugation operator~\cite{BD}, then, 
For our antiHermitian Yukawa model \eqref{ee1}, we  obtain, 
\begin{align}\label{resultcpt} 
&\widehat{\theta_{{\mathbb C}\mathcal{PT}}} \, \overline \psi (x) \, \gamma^5 \, \psi (x) \widehat{\theta_{{\mathbb C}\mathcal{PT}}}^{-1} = - \overline \psi (-x) \, \gamma^5 \, \psi (-x) \, \nonumber \\ 
&\Rightarrow \, \int d^4 x \Big[  \widehat{\theta_{{\mathbb C}\mathcal{PT}}}\, , \, 
\overline \psi (x) \, \gamma^5 \, \psi (x) \Big] |E\rangle = 2 \int d^4 x \, \widehat{\theta_{{\mathbb C}\mathcal{PT}}}\, \overline \psi (-x) \, \gamma^5 \, \psi (-x) |E\rangle = 0\, 
\end{align}
where in the last equality we took into account  \eqref{mueigen}, and the interpretation  that the mass eigenvalues of the Hamiltonian operator are associated with the dynamically generated masses for the various fields (fermions and axions) in the system. 

The result \eqref{resultcpt} implies that the non-Hermitian Yukawa interactions do {\it not affect} the equality of the dynamically-generated masses between particle and antiparticle (fermion or boson) states, in this system. Thus, the SD treatment of \cite{R11,R15}, and also the NJL analysis in this work, provides a correct framework within our anti-Hermitian Yukawa framework for a description of dynamical mass generation for both fermions and (pseudo)scalars.
The calculation of the correct form of the $\Theta$ operator using such ideas as the similarity transformation remains to be done for  non-Hermitian theories, having antilinear symmetries such as \cPT.

\section{The Yukawa model \label{sec:yuk}}

The massive Yukawa model is given by the bare Lagrangian in $3$-space and $1$-time dimensions in terms of bare parameters (emphasised through the use of the subscript $0$):\footnote{Our Minkowski-metric signature convention is $(+,-,-,-)$. For our discussion of dynamical mass generation the $\phi^{3}$ coupling is not going to be considered any further since our original Yukawa model does not require it for a consistent perturbative renormalisation, i.e. $\tilde g$ is not generated through renormalisation.}  
\begin{equation}
\label{e1}
\mathcal{L}= \frac{1}{2} \partial_{\mu} \phi_0 \partial^{\mu} \phi_0 -
\frac{M_0^2}{2} \phi_0^2 + \bar{\psi}_0 \left( i \slashed{\partial} - m_0 \right)
\psi_0 - i g_0 \bar{\psi}_0 \gamma^5 \psi_0 \phi_0 + \frac{u_0}{4!} \phi_0^4.
\end{equation}
$\mathcal{L}$ is renormalised through mass, coupling constant and wavefunction renormalisations; we will take the spacetime dimensionality $D$  to be  $4-\epsilon$ where $\epsilon$ is a small parameter. Furthermore $\epsilon$ is a useful small parameter in the analysis of fixed points. It is the simplest non-trivial renormalisable
model of a Dirac fermion field $\psi_0$ interacting with a pseudoscalar field $\phi_0$. \nt{If $g_0$ is real then the Yukawa term is Hermitian and $g_{0}^{2}>0$. If $g_0$ is \ntt{purely imaginary, then the Yukawa term is non-Hermitian but it is \cPT-symmetric, with our definitions of the discrete symmetries $\mathcal P, \mathcal T$ to be discussed below ({\it cf.} \eqref{pandt} and \eqref{phipt}), and $g_{0}^{2}<0$.} $u_0$ is real but it can be positive or negative. If $u_{0}>0$ the quartic term is non-Hermitian (in a Minkowski formulation). If $u_{0}<0$ the quartic term is Hermitian. Thus both couplings allow the possibility  of showing non-Hermitian but  \cPT-symmetric behaviour.} For the non-Hermitian case for $u$ (in the Euclidean picture) the path integral contour in the $\phi$ plane has a pair of \cPT symmetric Stokes wedges (see Fig.~\ref{fig:diagQuartic}) which means that the contour asymptotically needs to end up in these wedges. Moreover in this non-Hermitian case there are three $\phi$-saddle points (or configurations such as bounces depending on $D$); the fluctuations around the trivial saddle point give standard perturbation theory and Feynman rules. The non-trivial saddle points give rise to (non-perturbative) instanton-like contributions.  So renormalisation flows near Gaussian fixed points and quartic Hermitian fixed points using Feynman rules should be a reasonable indicator of the scale dependence of couplings of the theory.

\nt{In the Dirac representation of gamma matrices, the conventional discrete  transformations on $\psi_0$ that we use~\cite{BD} are:\footnote{\nt{We remark that in ref.~\cite{R13} a rather different $\mathcal{T}$ transformation was used related to the discrete symmetries of the Dirac equation \cite{BD}); the action of $\mathcal{T}$ on a spinor wave function $\psi$ produces the complex conjugate field $i \gamma^1\,\gamma^3\psi^\star (-t,\vec x)$. Such a transformation is a symmetry of the Lagrangian.\ntt{The form of these transformations in the Weyl representation of the $\gamma$ matrices are discussed in~\cite{schwartz}. The Lagrangian is invariant under these transformations}} }
\begin{align}\label{pandt}
\mathcal{P}\psi_0 (t,\vec{x})\mathcal{P^{\it{-1}}}=\gamma^{0}\psi_0 (t,-\vec{x}), \quad \mathcal{T}\psi_0 (t,\vec{x})\mathcal{T^{\it{-1}}}=i\gamma^{1}\gamma^{3}\psi_0(-t,\vec{x}), \quad {\mathbb C} \psi \left(t,\overrightarrow{x}\right) {\mathbb C}^{-1}=i\gamma^{2}\psi^{\dagger}\left(t,\overrightarrow{x}\right).
\end{align}
\ntt{where $\mathbb C$ denotes the  charge conjugation operator~\cite{BD} ({\it not} to be confused with the $\mathcal C$-operator) and $\mathcal{T}$ is the antilinear operator time-reversal operator.} \ntt{Also, under the action of $\mathcal P$ and $\mathcal T$, the pseudoscalar field $\phi\left(t,\overrightarrow{x}\right)$ transforms
as \footnote{\ntt{We note that, in \emph{quantum mechanics}~\cite{R1}, under $\mathcal T$ one has $i \to -i$, as a consequence of the action of $\mathcal T$ on the Heisenberg commutator between position ($\widehat x$) 
and momentum ($\widehat p$) operators: $\mathcal T\, [ \widehat x, \widehat p ] \, \mathcal T^{-1} = 
- [ \widehat x, \widehat p ] = - i $ ( $\hbar =1$ in natural units ), from which it follows immediately  that 
\be
\label{heiscomm}
\mathcal T\, i \, \mathcal T^{-1}  = -i~.
\ee
In canonical-quantisation of field theory~\cite{BD,bjr}, the above  Heisenberg-commutator argument, is extended to equal-time canonical commutators between fields and their canonical conjugate momenta, and, thus, the property \eqref{heiscomm} is understood to be valid for quantum field theoretic systems as well, and should be imposed when considering time-reversal transformations in the field theory lagrangian.
}}
\begin{align}\label{phipt}
\mathcal{P}\phi_0 (t,\vec{x})\mathcal{P^{\it{-1}}}= -\phi_0 (t,-\vec{x}), \qquad   \mathcal{T}\phi_0 (t,\vec{x})\mathcal{T^{\it{-1}}}= \phi_0 (-t,\vec{x}).
\end{align}}}

In this article we are interested in determining the conditions under which there is dynamical-mass generation. We stress that the results of our study here (and also those in \cite{R15a,R15,R11}) demonstrate the possibility of generating dynamically {\it real masses} in non-Hermitian theories. 
We use Feynman rules (discussed earlier in this article) at a perturbative level, which are valid for weak pseudo-Hermitian interactions. 

\ntt{We now come to the pseudoscalar self-interaction term in \eqref{e1}. For $u, \delta>0$, in any space dimension, the non-Hermitian scalar potential  
\begin{align}\label{ptnh}
u \, \phi^{2}{(i\phi)}^{\delta}~,
\end{align}
is \cPT-symmetric.}

For $\delta \neq 0$ the choice of $\mathcal T$ odd  for $\phi$ would spoil \cPT symmetry, and so is not of interest.
We have seen in our earlier discussion of Stokes wedges that the limit $\delta \to 2$, where $ u =u_0$, gives a \emph{non-Hermitian} but \cPT-symmetric theory; of course the term \eqref{ptnh} is 
non-Hermitian for every $4>\delta > 0$ and $u >0$. \cPT-symmetric Hamiltonians are pseudo-Hermitian. 

Explicitly the Lagrangian interaction term \eqref{ptnh} is pseudo-Hermitian and can be  derived by a similarity transformation from the Hermitian 
interaction $\phi^{2+\delta}$~\cite{antilin}:\footnote{In fact, exploiting the time-independence of the respective Hamiltonian, it suffices to evaluate the similarity transformation only for $t=0$.}
\begin{align}\label{simil}
- \, \phi (t=0, \vec x)^{2}{(i\phi (t=0, \vec x))}^{\delta} = S\, \phi(t=0, \vec x)^{2 + \delta}  \, S^{-1}, \quad S=\exp\Big(- \, \frac{\pi}{2} \, \int d^3x \, \Pi (t=0, \vec x) \, \phi(t=0, \vec x)\Big)~,
\end{align} 
where $\Pi(t, \vec x)$ is the canonical momentum of $\phi(t, \vec x)$, in the free theory, satisfying 
the (equal-time) canonical commutation relations $\Big[ \phi (t, \vec x)\, , \Pi (t, \vec x^\prime)\Big]=i \, \delta^{(3)}(\vec x - \vec x^\prime)$.\footnote{In arriving at \eqref{simil}, we use the Baker-Hausdorff foirmula 
$$ e^A \, B \, e^{-A} = B + \Big[A\, , \, B\Big] + \sum_{n=2}^\infty \frac{1}{n!} \Big[ A\, , \Big[ A \, , \dots \Big[A \, , \, B\Big] , \dots \Big], $$
and took into account the following result of the canonical field-theoretic (equal-time) commutation relation:  $$- \frac{\pi}{2}\, \Big[ \int d^3 x \, \Pi(t=0, \vec x)\, \phi(t=0, \vec x) \, , \, \phi(t=0, \vec y)^{2+\delta} \Big]= 
+i\, \frac{\pi}{2} \, (2 + \delta) \, \phi(t=0, \vec y)^{2+\delta}, \quad \delta > 0~,$$
which stems from the fact that $\Big[\, \, , \, \, \Big]$ is a linear operator, that behaves like a derivative with respect to the field $\phi$.}

In this work we will study the renormalisation of the Yukawa theory \eqref{e1} and determine its fixed point structure and the corresponding stability of the fixed points. We shall discuss RG flows that interpolate between Hermitian and non-Hermitian fixed points,  and discuss mass generation for the fermion and axion fields using the method of Nambu and Jona-Lasinio~\cite{R16, R16a}.   Hence we shall study mass generation for both axions and fermion fields using mass renormalisations calculated perturbatively, and examine the effects of the self-interaction on mass generation, by comparing our results with some of the non-perturbative masses discussed in \cite{R15,R11}. This discussion will be a prelude to the full Schwinger-Dyson treatment, with the inclusion of axion self interactions, reserved for a future publication. 

The  renormalised Lagrangian  (where the renormalised parameters are without the subscript $0$) is given by 
\begin{eqnarray}
\mathcal{L}&=& \frac{1}{2} (1 + \delta Z_{\phi}) \partial_{\mu} \phi
\partial^{\mu} \phi - \frac{M_0^2}{2} (1 + \delta Z_{\phi}) \phi^2 + (1 +
\delta Z_{\psi}) \bar{\psi} \left( i \slashed{\partial} - m_0 \right) \psi \nonumber\\
& & - i g_0(1 + \delta Z_{\psi}) \sqrt{1 + \delta Z_{\phi}} \bar{\psi} \gamma^5 \psi \phi
+ \frac{u_0}{4!} (1 + \delta Z_{\phi})^2 \phi^4 
\label{e3a}
\end{eqnarray}
 where we have introduced the multiplicative renormalisations $Z_{\phi}$, $Z_{\psi}$, $Z_{g}$, $Z_{u}$, $Z_{m}$, and $Z_{M}$ defined through
\begin{eqnarray}
\phi_0 &=& \sqrt{ Z_{\phi}} \phi = \sqrt{1 + \delta Z_{\phi}} \phi,\label{e5}\\
\psi_0 &=& \sqrt{ Z_{\psi}} \psi= \sqrt{1 + \delta Z_{\psi}} \psi,\label{e6}\\
M^2_0 Z_{\phi} &=& M^2 + \delta M^2=M^2 Z_{M}, \label{e7}\\
m_0 Z_{\psi} &=& m + \delta m = m Z_{m}, \label{e8} \\
g_0 Z_{\psi} \sqrt{Z_{\phi}} &=& g + \delta g=g Z_{g}, \label{e9}\\
u_0 (Z_{\phi})^2 &=& u + \delta u=u Z_{u}.\label{e10}
\end{eqnarray}

This Yukawa model  is the natural field-theoretic version of the quantum mechanical model considered in~\cite{R13}, which can be considered as a free theory but with both conventional and axial mass terms. The axial mass becomes a dynamical field in our model within a Yukawa term; the Yukawa coupling $g$ can be  purely imaginary $g^2 < 0$ ~\cite{R15a,R11,R15}, due to the aforementioned \cPT symmetry of the relativistic theory~\cite{R15a,antilin}.\ntt{\footnote{\ntt{We note that purely imaginary couplings, upon renormalization, are consistent in our models, in the sense that the respective counterterms $Z_g = 1 + \mathcal O(g^2)$, $Z_\psi$ and $Z_\phi$ in \eqref{e9} are real.}}} It was noted in~\cite{R13} that the \emph{absence} of a conventional mass term led to \emph{broken}  \cPT  symmetry\footnote{\nt{A system with broken \cPT symmetry is one in which there are some energy eigenvalues which occur in complex conjugate pairs. A system with unbroken \cPT symmetry is one in which all the energy eigenvalues are real. } } and so it is natural, at the perturbative level, to consider a massive theory. Moreover one approach to dynamical mass generation~\cite{R16a}, the one that we will follow, is to consider a theory with a mass  which is then determined self-consistently through a gap equation~\cite{R16}.

The massless variant of the Yukawa theory ($M_{0} =m_{0}=0$), with no quartic term, has recently been studied  using unrenormalised Schwinger-Dyson equations~\cite{R15,R11} with a momentum cut-off $\Lambda$. The emphasis was on effective theory for energy and momentum scales below $\Lambda$. Earlier work has suggested a link between renormalisation and emergence of non-Hermiticity~\cite{R9}. Our fermionic model \eqref{ee1}, a natural generalisation of the canonical scalar model \eqref{E1}, shows the interplay of non-Hermiticity and \cPT symmetry. The associated renormalization group allows us to discuss the energy dependence of the couplings and the Hermiticity of the theory.  We should note that in \cite{R15,R11}  non-Hermiticity was examined only for the coupling $g$. 
Here, we examine \cPT-symmetric non-Hermiticity in the self-interaction of the pseudoscalar field \eqref{ptnh} as well.
Nonetheless, as we will find that the dynamically generated fermion masses will acquire a non-perturbative form, similar in structure to the one derived in \cite{R15,R11} for real coupling $g$.
\section{Renormalsation group analysis of the massive Yukawa theory}\label{MYukawa}

We shall use dimensional regularisation of the massive Yukawa theory with spacetime dimension $D=4-\epsilon$ and $\epsilon >0$. At one loop the renormalisation group equations are:
\begin{eqnarray}
\frac{dg}{dt} & = & \frac{5g^{3}}{16\pi^{2}}-\frac{\epsilon g}{2} \label{e11} \\
\frac{du}{dt} & = & \frac{48 g^{4}-3u^{2}+8g^{2}u}{16\pi^{2}} -u\epsilon\label{e12}\\
\frac{dm}{dt} & = &-\frac{g^{2}m^{2}}{16\pi^{2}}\label{e13}\\
\frac{dM}{dt} & = &\frac{1}{32\pi^{2}M}\left[4g^{2} (M^{2}-2m^{2}) -um^{2}   \right] \label{e14}
\end{eqnarray}
where $\frac{d}{dt}\equiv\mu\frac{d}{d\mu}$, $\mu$ being the mass scale introduced in the method of dimensional regularisation. The study of $\epsilon$-dependent fixed points was initiated by Wilson and Fisher~\cite{R17}.

It is also interesting to consider a change of variables from $(m,M,g)$ to $(\sigma, M, y)$ where $m=\sigma M$  and $y=g^2$ in order to see any correlation in the behaviour of $m$ and $M$ under renormalisation. The $\beta$-functions are then polynomials in these variables.From \eq{e11}, \eq{e12}, \eq{e13} and \eq{e14} we deduce that 
\begin{eqnarray}
\frac{dM}{dt} & = & \frac{M}{32\pi^{2}}\left[4 y(1-2\sigma^{2})-\sigma^{2}u\right] \label{e15}\\
\frac{d\sigma}{dt} & = & -\frac{\sigma}{32\pi} \left[4y(1-2\sigma^{2}(1+\frac{M}{128\pi^{}})) -u\sigma^{2}  \right]\label{e16}\\
\frac{d y}{dt}& = &\frac{5}{8\pi^{2}}y^{2}-\epsilon y \label{e16a}\\
\frac{d u}{dt}& = &\frac{48 y^{2}-3u^{2}+8yu}{16\pi^{2}} -u\epsilon\label{e16b}
\end{eqnarray}
Equations \eq{e11} and \eq{e12} form a closed set; their solutions feed into the equations \eq{e13} and \eq{e14}. Similarly \eq{e16a} and \eq{e16b} form a closed set.

Our strategy will be to use a combination of dominant-balance ideas for equations~\cite{R20}, geometric methods from the theory of dynamical systems and direct solution of differential equations to determine flows to the Hermitian and non-Hermitian regions of parameter space. Our conclusions will be valid within the context of the above (approximate) one-loop renormalization group equations. 

In the next section we shall analyse the two-loop renormalization group flows for a massless Yukawa theory, a model that was  previously used in a Schwinger-Dyson  analysis of mass generation \cite{R11, R15}.

\subsection{The behaviour of the $g$ and $u$ coupling constants}

The fixed points of $g$ are $g^*$  where  $g^*=g^*_{\pm}=\pm \sqrt{\frac{8 \pi^{2}\epsilon}{5}}$ and the trivial fixed point $g^*=0$.\footnote{The sign of $g$ distinguishes separate parts of "theory" space. } The related fixed points $u^*$ for $u$ are determined by
\begin{equation}
48{g^{*}}^{4}-3{u^{*}}^{2}+8{{g^{*}}^{2}}u^{*}=16\pi^{2}\epsilon {u^{*}}.
\label{e17}
\end{equation} 

The solutions for $u^{*}$  are $u^{*}=0$ and $u^{*}_{\pm}=u_{\pm}\epsilon$ where

\begin{equation}
\label{e18}
u_{\pm}=\frac{8}{3}g_{0}^{2} \pm \sqrt{\frac{64}{9}g_{0}^{4}+64}
\end{equation}
and $g_{0}=\sqrt{\frac{8\pi^{2}}{5}}\sim 3.97$, which gives $u_{+}\sim 84.97$ and $u_{-}\sim -0.75$. Thus, we 
observe that $u_{-}$ is negative and, therefore according to our discussion below \eqref{ptnh} in section \ref{sec:yuk}, is a Hermitian fixed point. On the other hand,  $u_{+}$ is positive and, thus, is a non-Hermitian fixed point.\footnote{In another  Hermitian model, to be discussed later, the possibility of flow to a non-Hermitian quartic self-coupling fixed point has also  been noticed~\cite{R20a}.} At the level of fixed points, non-Hermiticity is therefore introduced through the $u$ coupling. On the other hand, $g$ remains real at the fixed points. In summary, the various fixed points  (in the $(g, u)$ plane), denoted by $(g^{*},u^{*})$, are given below: 
\begin{enumerate}
  \item (0,0)
    \item (0, -$\frac{16\pi^{2}}{3}\epsilon$)
  \item ($g^{*}_{+},u^{*}_{+}$)
    \item ($g^{*}_{+},u^{*}_{-}$)
      \item ($g^{*}_{-},u^{*}_{+}$)
        \item ($g^{*}_{-},u^{*}_{-}$)
\end{enumerate} 
which will be denoted by $f_{i}, i=1,\cdots,6$.~The $\epsilon$-dependent fixed points are examples of Wilson-Fisher fixed points \cite{R17}. Let us first discuss the linear stability of these fixed points. 

\subsubsection{Stability of fixed points in the ($u$,$g$) plane\label{sec:stability}}

We denote deviations from the fixed points by $\delta g=g-g^*$ and $\delta u=u-u^*$. Linear stability analysis around the fixed point gives
\begin{equation}
\label{e19 }
\frac{d}{dt}\left(\begin{array}{c}\delta g \\\delta u\end{array}\right)=M(g^*,u^*,\epsilon)\left(\begin{array}{c}\delta g \\\delta u\end{array}\right).
\end{equation}

In order to avoid algebraic complexity we will consider the stability using a numerical value for $\epsilon=.0101321$ which leads to $g^*_{\pm}=\pm.4$, $u^*_{+}=.887953$ and $u^*_{-}=-.461286$. Eigenvalues for $M$ at the fixed point $f_i$ are $\lambda_{i1}$ and $\lambda_{i2}$. The corresponding (unnormalised) two-dimensional 
 eigenvectors are $e_{i1}$ and $e_{i2}$. Thus, we have:

 \begin{enumerate}
  \item  $\lambda_{11}=-.0101321$ and  $\lambda_{12}=-.00506605$ with $e_{11}=\left(\begin{array}{c} 0 \\1\end{array}\right)$ and  $e_{12}=\left(\begin{array}{c} 1 \\0\end{array}\right)$,
  \item $\lambda_{21}=.0101321$ and  $\lambda_{12}=-.00506605$ with $e_{21}=\left(\begin{array}{c} 0 \\1\end{array}\right)$ and  $e_{22}=\left(\begin{array}{c} 1 \\0\end{array}\right)$,

  \item   $\lambda_{31}=-.0357646$  and $\lambda_{32}=.0101321$ with $e_{31}=\left(\begin{array}{c} 0 \\1\end{array}\right)$ and  $e_{32}=\left(\begin{array}{c} .37403 \\.927417\end{array}\right)$,

  \item   $\lambda_{41}=.0155004$   and $\lambda_{42}=.0101321$ with $e_{41}=\left(\begin{array}{c} .0 \\1\end{array}\right)$ and  $e_{42}=\left(\begin{array}{c} .0904311 \\-.995903\end{array}\right)$,

  \item   $\lambda_{51}=-.0357646$  and $\lambda_{52}=.0101321$  with $e_{51}=\left(\begin{array}{c} 0 \\1\end{array}\right)$ and  $e_{52}=\left(\begin{array}{c} .37403 \\-.927417\end{array}\right)$,
  
  \item   $\lambda_{61}=.0155004$  and $\lambda_{62}=.0101321$ with $e_{61}=\left(\begin{array}{c} 0 \\1\end{array}\right)$ and  $e_{62}=\left(\begin{array}{c} 0.0904311 \\.995903\end{array}\right)$.

\end{enumerate}

A fixed point $f_{i}$ is:
 \begin{itemize}
  \item a sink if  $\lambda_{i1}<0$ and $\lambda_{i2}<0$
  \item a source if  $\lambda_{i1}>0$ and $\lambda_{i2}>0$
  \item a saddle point if  $\lambda_{i1}>0$ and $\lambda_{i2}<0$ or vice versa.
\end{itemize}

These fixed points help to organise the renormalisation group flow through their basins of attraction. If an eigenvalue is $0$, then a nonlinear analysis is required around the fixed point to determine its stability. We note that $t \to \infty$ is a flow to high energy; the flow $t \to -\infty$ is a flow to low energy. An energy of $O(1)$ corresponds to $t=0$.We shall consider the fixed points for $m$ and $M$ later.

\subsubsection {The renormalisation group flow for $u$ and $g$\label{sec:group}}

We shall consider the solutions of the coupled flow equations  \eq{e11} and \eq{e12} (and the closely related equations \eq{e17} and \eq{e18}). We can rewrite \eq{e11} as
\begin{equation}
\label{e20}
\frac{dg}{dt}=\frac{5g}{16\pi^{2}}(g-g_{+})(g-g_{-}).
\end{equation}

For $g\gg g_{+}$ \eq{e20} simplifies to
\begin{equation}
\label{e21}
\frac{dg}{dt}=\frac{5}{16\pi^{2}}g^3
\end{equation}
and leads to 
\begin{equation}
\label{e22}
g^2=y=-\frac{1}{2(c+\frac{5t}{16\pi^{2}})}
\end{equation}
where  $c$ is a constant of integration. At $t=0$, if the theory is Hermitian, then $c$ is negative. As $t$ increases, $g$ increases  but remains Hermitian until at finite time $t=\frac{16\pi^{2}|c|}{5}$ the approximation of small $g$, and thus perturbative renormalisation, breaks down. 

For $g\ll{g_{-}}$ we again  have \eq{e21} and $c$ is negative for a theory which is Hermitian at a scale $\mu \sim 1$. In the IR, $g$ remains small. In the UV, $g$ moves towards $g=0$ but then veers  away to large positive values of $g$ where perturbation theory is not trustworthy.

For $0<g<g_{+}$ it is clear that $g  \to 0$ as $t \to \infty$. As $t \to -\infty$ we have $g \to g_{+}$. As $\epsilon \to 0$ there is a bifurcation where the fixed points $g_{+}, g_{-}, 0$ coalesce. The trivial fixed point is unstable both in the IR and the UV.

We will now consider the flow of $u$ using \eq{e16b}. The solution of \eq{e16a} is 
\begin{equation}
\label{e33}
y(t)=-\frac{8d(\epsilon)\pi^{2}\epsilon}{5(e^{\epsilon t}-d(\epsilon)}
\end{equation}
where $d(\epsilon)=e^{8\pi^{2}\epsilon c_{1}}(>0)$ and $c_{1}$ is a constant of integration. The resultant solution of \eq{e16b} for $u(t)$ is 
\begin{equation}
\label{e34}
u(t)=8 d(\epsilon){\pi^{2}}\epsilon (1+\sqrt{145}-\frac{c_{2}(\sqrt{145}-1)e^{\sqrt{29/5}\epsilon t}}{(e^{\epsilon t}-d(\epsilon))^{\sqrt{29/5}})})/({15(e^{\epsilon t}-d(\epsilon)) (1+\frac{c_{2}e^{\sqrt{\frac{29}{5}}\epsilon t}}{(e^{\epsilon t}-d(\epsilon))^{\sqrt{29/5}}})})
\end{equation}
where $c_2$ is an integration constant.
This is complicated to analyse. If we keep away from the region of the fixed points near the origin, by considering  $\epsilon \to 0$, the solutions in \eq{e33} and \eq{e34} can be simplified to
\begin{equation}
\label{e35}
y(t)=-\frac{1}{\frac{5}{8\pi^{2}} t+c_{1}}
\end{equation}
and
\begin{equation}
\label{e36}
u(t)=\frac{8\pi^{2}\left[1-\sqrt{145}+\left(1+\sqrt{145}\right)\left(8c_{1}\pi^{2}+5t\right)^{\sqrt{\frac{29}{5}}}c_{2}\right]}{3(8c_{1}\pi^{2}+5t)(1+c_{2}(8c_{1}\pi^{2}+5t)^{\sqrt{29/5}})}.
 \end{equation}
For $g$ to be non-Hermitian at $t=0$, we have $y<0$, and so $c_{1}>0$; so as $t \to \infty$, $y$ remains non-Hermitian but slowly vanishes. In the infrared (IR), as $t$ decreases from $t=0$, $y$ increases but remains non-Hermitian; perturbation theory becomes unreliable.

In the Hermitian case, $y>0$ at $t=0$ and so $c_1$ is negative. As $t$ increases from $t=0$ $y$ remains Hermitian but increases until perturbation theory is invalid.

For $u$ to be real we need $(8 c_{1}\pi^{2}+5 t)$ to be nonnegative. This requires $c_2=0$ and so 
\begin{equation}
\label{e37}
u(t)=-\frac{8\pi^{2}\left(\sqrt{145}-1\right)}{3\left(8c_{1}\pi^{2}+5t\right)}.
 \end{equation}
 
 The implication of a non-Hermitian $g$ ($c_{1}>0$) for $u$ is that it is Hermitian ( i.e. $u<0$ ) at $t=0$. As $t \to \infty$ , $u$ falls-off to $0$ but remains Hermitian. In the IR, $u$ increases until perturbation theory is unreliable.  
 
  The implication of a Hermitian $g$ ($c_{1}<0$) is that $u$ is non-Hermitian ( i.e. $u>0$ ) and remains so in the IR. The self-interaction coupling $u$ increases in the UV  until the perturbative analysis becomes unreliable. In the IR, $u$ falls-off but remains non-Hermitian.

\subsubsection{The renormalisation group flow for $m$ and $M$\label{sec:flows}}

The fixed points for $M$ and $\sigma$ can be deduced~\footnote{In principle the poles of Green's function and $m$ and $M$ are distinct.} from \eq{e15} and  \eq{e16}. Let ( $u^*$ ,$y^*$ ) denote any of the fixed points that we have already  found. The possible fixed points are:
\begin{itemize}
  \item $(M^*=0,\,\sigma^*=0)$ 
  \item $(M^*=0,\,l(y^*,\sigma^*)=0)$  
  \item $(l(y^*,\sigma^*)=0, \sigma^*=0)$
\end{itemize}
where $l(y,\sigma)=4y(1-2\sigma^2)-u\sigma^2$. On analysing these possibilities, we find the fixed points are $\sigma^*=M^*=0$ in addition to the fixed points for $u$ and  $g$.

 The  beta functions associated with \eq{e15}, \eq{e16}, \eq{e16a} and \eq{e16b} are:
\begin{eqnarray}
\beta_{y}(y,\epsilon) & = & \frac{5y^{2}}{8\pi^{2}}-\epsilon y \\
\beta_{u}(y,u,\epsilon) & = & \frac{48 y^{2}-3u^{2}+8yu}{16\pi^{2}} -u\epsilon\\
\beta_{\sigma}(y,u,\sigma,M) & = &-\frac{\sigma}{32\pi} \left[4y(1-2\sigma^{2}(1+\frac{M}{128\pi^{}})) -u\sigma^{2}  \right]\\
\beta_{M}(y,u,\sigma,M) & = &\frac{M}{32\pi^{2}}\left[4y (1-2\sigma^{2})-\sigma^{2}u\right]
 \end{eqnarray}

In ($y,u,\sigma,M$)-space consider $\delta y \equiv y-y^*$, $\delta u \equiv u-u^*$, $\delta \sigma \equiv \sigma-\sigma^*$ and $\delta M \equiv M-M^*$. These linear deviations around a generic fixed point $(y^*,u^*,\sigma^*,M^*)$ satisfy
\begin{equation}
\label{RG1fp}
\frac{d}{dt}\left(\begin{array}{c}
\delta y\\
\delta u\\
\delta\sigma\\
\delta M
\end{array}\right)=\underline{\underline{N}}\left(\begin{array}{c}
\delta y\\
\delta u\\
\delta\sigma\\
\delta M
\end{array}\right)
 \end{equation}
where $\underline{\underline{N}}$ is a $4\times 4$ matrix whose non-zero elements are given by
\begin{eqnarray}
N_{11}& = &\partial_{y}\beta_{y}\left(y^{*},\epsilon\right) \, =
\frac{5 y^*}{4 \pi ^2}-\epsilon \\
N_{21}& = &\partial_{y}\beta_{u}\left(y^{*},u^{*},\epsilon\right)\,=\frac{8 u^*+96 y^*}{16 \pi ^2} \\
N_{22}& = &\partial_{u}\beta_{u}\left(y^{*},u^{*},\epsilon\right) \,= \frac{8 y^*-6 u^*}{16 \pi
   ^2}-\epsilon   \\
N_{31}& = &\partial_{y}\beta_{\sigma}\left(y^{*},u^{*},\sigma^{*},M^{*}\right) \,=-\frac{M^* {{\sigma}^{*} }^{2}}{16 \pi
   ^2}+\frac{{\sigma^*} ^3}{4 \pi
   ^2}-\frac{{\sigma}^* }{8 \pi
   ^2}\\
N_{32}& = &\partial_{u}\beta_{\sigma}\left(y^{*},u^{*},\sigma^{*},M^{*}\right)\,=\frac{{\sigma ^*}^3}{32 \pi ^2} \\
N_{33}& = &\partial_{\sigma}\beta_{\sigma}\left(y^{*},u^{*},\sigma^{*},M^{*}\right)\,=-\frac{M^* \sigma^*  y^*}{8 \pi
   ^2}+\frac{3 {\sigma^* }^2
   \left(\frac{u^*}{8}+y^*\right)}
   {4 \pi ^2}-\frac{y^*}{8 \pi
   ^2} \\
N_{34}& = &\partial_{M}\beta_{\sigma}\left(y^{*},u^{*},\sigma^{*},M^{*}\right)\,=-\frac{{\sigma^* }^2 y^*}{16 \pi ^2} \\
N_{41}& = &\partial_{y}\beta_{M}\left(y^{*},u^{*},\sigma^{*},M^{*}\right) \,=\frac{M^* \left(1-2 {\sigma^*}
   ^2\right)}{8 \pi ^2}\\
N_{42}& = &\partial_{u}\beta_{M}\left(y^{*},u^{*},\sigma^{*},M^{*}\right)\,=-\frac{M^* {\sigma^*} ^2}{32 \pi ^2} \\
N_{43}& = &\partial_{\sigma}\beta_{M}\left(y^{*},u^{*},\sigma^{*},M^{*}\right)\,=-\frac{M^* ( \sigma^*  u^*+8
   \sigma^*  y^*)}{16 \pi ^2} \\
N_{44}& = &\partial_{M}\beta_{M}\left(y^{*},u^{*},\sigma^{*},M^{*}\right)\,=\frac{4 \left(1-2 {\sigma^*}
   ^2\right) y^*-{\sigma^*} ^2 u^*}{32
   \pi ^2}
\end{eqnarray}

The fixed points in the space $(y, u, \sigma, M)$ are 
\begin{itemize}
  \item $f_{7}=(0,0,0,0)$
  \item $f_{8}=(0,-\frac{\epsilon}{2},0,0)$
  \item $f_{9}=(y_{+}^{*},u_{+}^{*},0,0)$
  \item $f_{10}=(y_{+}^{*},u_{-}^{*},0,0)$
\end{itemize}

Recall the values of the fixed points used earlier in the $(g,u)$ plane:  ${y_{+}}^*={{g_{+}}^*}^{2}=0.16$, ${u_{+}}^*=0.887953$, and ${u_{-}}^*=-0.461286$. In terms of $m$ and $M$ the fixed points are $m=m^*=0$ and $M=M^*=0$. From \eq{e13} and \eq{e14} we note that near these fixed points, for non-trivial $g$,
\begin{eqnarray}
\frac{dm}{dt}& = &-\frac{g^{*2}}{8\pi^{2}}m, \label{e38}\\
\frac{dM}{dt}& = &\frac{g^{*2}}{8\pi^{2}}M \label{e39}.
\end{eqnarray}

Since \eq{e38} and \eq{e39} do not depend on $u^*$, they are not affected by non-Hermiticity. We note that:
\begin{itemize}
  \item  A small deviation of $m$ increases (decreases) in the IR (UV). 
  \item   A  small deviation of $M$ increases (decreases) in the UV (IR).
\end{itemize}

\section{Dynamical mass generation \label{sec:NJL}}

A non-perturbative approach to dynamical mass generation was pioneered by Nambu and Jona-Lasinio (NJL)~\cite{R16a}, extending the mechanism for the generation of a mass gap in superconductivity to relativistic particle physics. In their model, NJL discussed dynamical chiral symmetry breaking via the generation of fermion masses through appropriate bilinear fermion condensates that were formed as a result of attractive (\emph{non-renormalisable}) four-fermion contact interactions. They restricted their discussion to one loop, and found that fermion mass generation was possible when the coupling of the four-fermion interactions exceeded a critical value.

In our approach, we shall use the model \eqref{e1}, which,  in contrast to the NJL model~\cite{R16a}, is renormalisable 
and does not contain four-fermion interactions. It contains, however, Yukawa interactions and pseudoscalar self-interactions. As shown below, these interactions suffice to generate dynamical masses for both fermion and pseudoscalar fields, for small values of the Hermitian (real) coupling of the  Yukawa interaction. \ntt{We shall make use of the arguments of \cite{R15a}, according to which the dynamical generation of a non-Hermitian chiral-fermion-mass term $m_5 \overline \psi \gamma_5 \psi$ is not energetically favourable, to discuss only dynamical generation of a Dirac-type mass for the fermion $\psi$ and a mass for the pseudoscalar (axion-like) field $\phi$.}

\begin{figure}[!h]
\centering\includegraphics[width=4.5in]{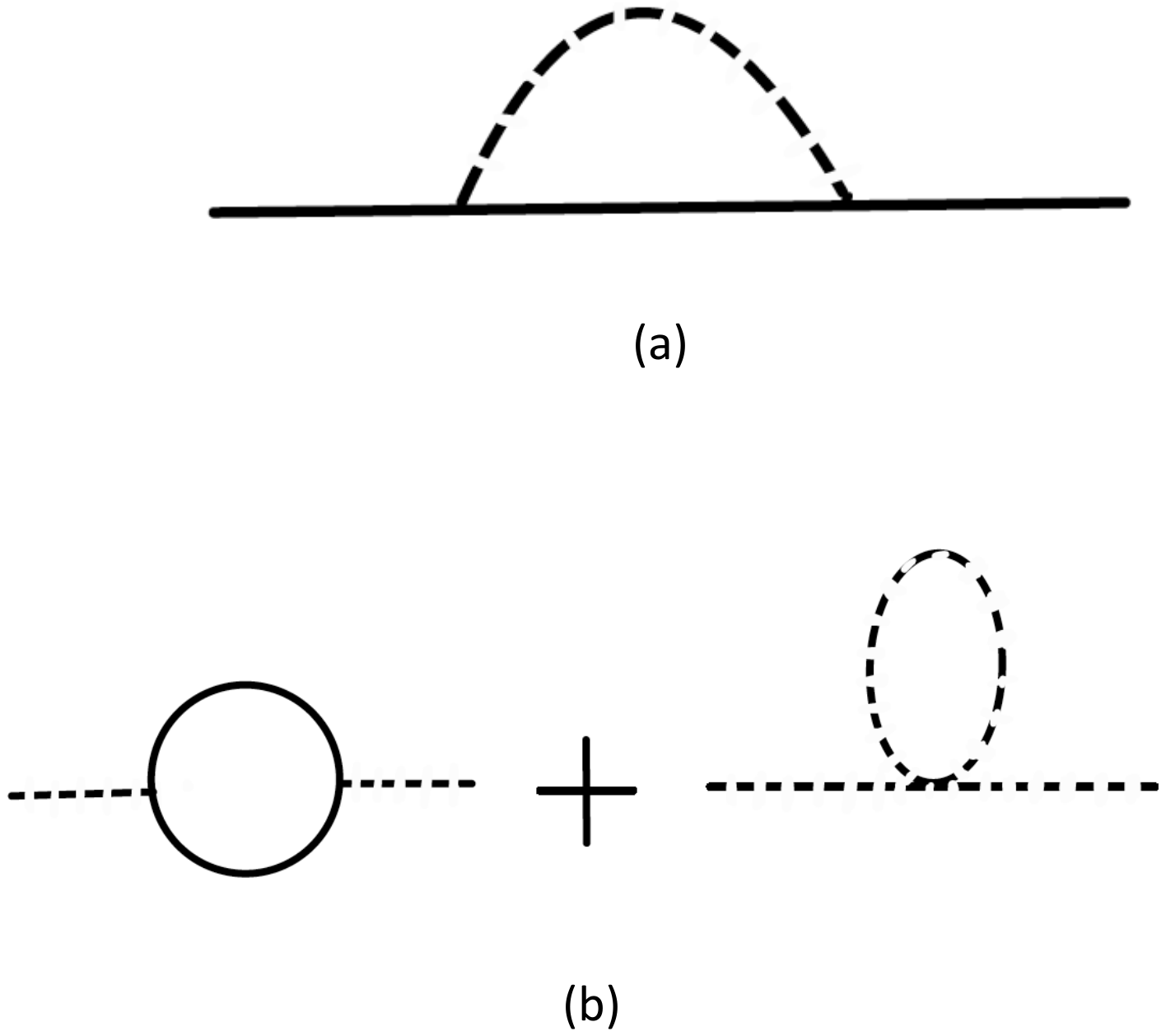}
\vspace{-6.5cm}
\caption{(a) One-Loop diagram contributing to the quantum corrections of the fermion self-energy in the model \eqref{e1}. Dashed lines correspond to pseudoscalar fields, whilst continuous lines indicate fermions. (b) One-loop diagrams contributing to the quantum corrections of the pseudoscalar self-energy.}
\label{fig:diag}
\end{figure}

We shall use the simplified (one-loop)~\cite{R16} NJL approach to check the possibility of dynamical mass generalisation in our model \eqref{e1}. Within the context of our renormalisable theory, the NJL  approach starts with a Lagrangian with nonzero renormalised masses $m$ and $M$~\cite{R16}. We will denote the massless free Lagrangian by $L_0$ (which contains the kinetic terms for the $\phi$ and $\psi$ fields)  and the interaction Lagrangian by $L_{int}$ (which contains the Yukawa and pseudoscalar self-interaction terms); thus the full  Lagrangian $L$ is
\begin{equation}
\label{e40}
(L_{0}-m\overline{\psi}\psi-\frac{1}{2}M^{2}\phi^{2})+(L_{int}+\varDelta m\, \overline{\psi}\, \psi+\frac{1}{2}\varDelta M^{2}\phi^{2}).
\end{equation}
At the end of the calculation of the two-point one-particle-irreducible (1PI) functions for the scalar and the fermion, we set $M^{2}=\varDelta M^{2}$ and $\varDelta m=m$~\cite{R16}.
The renormalised two point 1PI functions for the fermion and scalar   are \emph{assumed} to behave like
\begin{eqnarray}
\varGamma_{f}^{\left(2\right)}\left(p\right)& = &\tilde{Z_{f}}\left(\gamma^{\nu}p_{\nu}-m\right) \\
\varGamma_{s}^{\left(2\right)}\left(p\right)& = &\tilde Z_{s}\left(p^{2}-M^{2}\right)
\end{eqnarray}
for $p^2 \ll m^2, M^2$, where $\tilde Z_{f}$ and $\tilde Z_{s}$ are finite renormalisations. From renormalised \emph{one-loop} perturbation theory for the fermion two-point function we can readily show that~\footnote{One loop involving two Yukawa vertices contributes to the fermion self-energy (see fig.~\ref{fig:diag}(a)).}
\begin{equation}
\label{e41}
m+\frac{g^{2}m}{16\pi^{2}}\int_{0}^{1}dx\left(\gamma+\log\left(\frac{\Delta\left(x,m,M\right)}{4\pi\mu^{2}}\right)\right)=0
\end{equation}
where $\gamma$ is the Euler constant and
\begin{equation}
\label{e42}
\Delta\left(x,m,M\right)=xm^{2}+\left(1-x\right)M^{2}.
\end{equation}

From renormalised \emph{one-loop} perturbation theory for the scalar two point function we can also show that~\footnote{Two Feynman diagrams contribute to the (pseudo)scalar self energy, one involving a fermion loop and the other a (pseudo)scalar loop (see fig.~\ref{fig:diag}(b)).}

\begin{equation}
 M^{2}+\frac{uM^{2}}{2}\left(\frac{\gamma-1+\log\frac{M^{2}}{4\pi\mu^{2}}}{16\pi^{2}}\right)+
\frac{g^{2}}{4\pi^{2}}\left[m^{2}(\gamma-1+\,\log\left(\frac{m^2}{4\pi\mu^{2}}\right))\right]=0. \label{e43}
\end{equation}

 This approach~\cite{R16} to  dynamical mass generation is approximate and relies on perturbative renormalizability. In order to analyse \eqref{e42} and \eqref{e43}  it is convenient first to introduce dimensionless variables 
 \begin{align}\label{variables}
 a=\frac{\mathit{m}}{2\sqrt{\pi}\mu} \quad {\rm and} \quad
 b=\frac{M^{2}}{4\pi\mu^{2}}.
 \end{align} 
 In terms of $a$ and $b$,  \eqref{e42} and \eqref{e43} read
 \begin{equation}
\label{e43a}
1=-\frac{g^{2}}{16\pi^{2}}\int_{0}^{1}dx\,\left(\gamma+\log\left[xa^{2}+\left(1-x\right)b\right]\right)\end{equation}
and
\begin{equation}
\label{e43b}
b+\frac{u\,b}{32\pi^{2}}\left(\gamma-1+\log b\right)+\frac{g^{2}a^{2}}{4\pi^{2}}\left(\gamma-1+\log a^{2}\right)=0
\end{equation}
respectively.
It is straightforward to show that 
\begin{equation*}
\int_{0}^{1}dx\,\log\left[xa^{2}+\left(1-x\right)b\right]=\frac{1}{a^{2}-b}\left[a^{2}\log a^{2}-a^{2}-b\log b+b\right]
\end{equation*}
which has a vanishing limit as $a^2 \to b$ ( a consistency requirement). 
For notational simplification we let 
\begin{align}\label{gubar}
\overline{g}^{2}\equiv\frac{g^{2}}{4\pi^{2}} \quad {\rm and} \quad \overline{u}\equiv\frac{u}{32\pi^{2}}. 
\end{align}
Then, Eq.~(\ref{e43a}) becomes
\begin{equation}
\label{e43c}
\left(1+\frac{\overline{g}^{2}\gamma}{4}-\frac{\overline{g}^{2}}{4}\left(1-\log a^{2}\right)\right)a^{2}=b\left(\frac{\overline{g}^{2}}{4}\left(\log b-1\right)+1+\frac{\overline{g}^{2}\gamma}{4}\right)
\end{equation}
and \eqref{e43b} becomes
\begin{equation}
\label{e43bcc}
\bar{u}b\log b=-\left(-\bar{u}\left(1-\gamma\right)+1\right)b-\bar{g}^{2}a^{2}\left(\gamma-1+\log a^{2}\right).
\end{equation}
We shall study the possible solutions of (\ref{e43c}) and (\ref{e43bcc}) \emph{in various limits}. Since our fixed points for $u$ and $g$ have been found in perturbation theory, we will not strictly adhere to their values at fixed points in considering the landscape of regimes where dynamical mass generation may be possible. 
 This landscape will guide future nonperturbative studies using the Schwinger-Dyson equations, which will appear in a forthcoming publication. We will consider the following limiting cases:
\begin{itemize}
  \item 
 
If $a^2$ and $b$ are both small, then from (\ref{e43c}) we have approximately the leading behaviour
\begin{equation}
\label{e43d}
\frac{\bar{g}^{2}}{4}a^{2}\log a^{2}=\frac{\bar{g}^{2}}{4}b\log b
\end{equation}
which is certainly compatible with $a^{2}=b$. 

Assuming then that $a^2 \simeq b$, we observe from (\ref{e43bcc}) that  
the leading behaviour gives
\begin{equation}
\label{e43e}
\overline{u}b\log b=-\overline{g}^{2}a^{2}\log a^{2}
\end{equation}
which would imply that a solution with small $a^2$ and small $b$  is possible if 
\begin{align}\label{ug2}
\overline{u} \simeq -\overline{g}^{2}~.
\end{align} 
From \eqref{e43c} we can also deduce that
\begin{equation}
\label{e43cc}
b\approx\exp\left(-\frac{4}{\bar{g}^{2}}-\gamma\right).
\end{equation}
So $g$ and $u$ would both have to be Hermitian and small for $b\,(\simeq a^2)$ to remain small. 
It is assumed that $\overline{g}^{2}$ is positive to avoid getting a $b$ which is too large, and so we stay here within the 
Hermitian Yukawa interactions. Hence in this approach a \emph{solution with small $a^{2}$ and $b$} is only compatible with small $\overline{g}$ in the Hermitian case (We remind the reader that, according to our discussion below \eqref{ptnh} in section \ref{sec:yuk}, negative $u$ corresponds to the Hermitian theory).

Hence mass generation can take place with $\overline{u}$ small and negative, for small $\overline{g}^{2}$. A Wilson-Fisher point which is qualitatively similar is $\overline{g}^{2}=.4\,\epsilon$ and $\overline{u}=-.0783\,\epsilon$.

By substituting $a^2=b$ in \eqref{e43b} 
and using \eqref{e43cc}, we obtain  
\begin{equation}
\label{e46}
\overline{u}=- \overline{g}^{2}\, \frac{\overline{g}^{2} + 3}{\overline{g}^{2}+4},
\end{equation}
which corrects \eqref{ug2} with higher orders in the Yukawa coupling. 

Using the definitions \eqref{variables}, \eqref{gubar}, we then arrive at  the following expression for the dynamically generated fermion and axion masses, assumed approximately equal in magnitude:
\begin{align}\label{famasses}
m \simeq M = \tilde \mu \, \exp\Big(-\frac{8\pi^2}{g^2}\Big), \quad \tilde \mu \equiv \sqrt{4\pi} \, e^{-\gamma/2}\, \mu \, .
\end{align}
$\tilde \mu$ is the transmutation mass parameter redefined, in the standard way, to absorb the Euler's constant $\gamma$. Given that $g $ is perturbatively small in our analysis, the dynamically generated masses \eqref{famasses} are {\it nonperturbative} in the real Yukawa coupling $g$.

It must be noted that the form of the solution \eqref{famasses} is {\it identical} to the one generated through a Schwinger-Dyson approach in \cite{R15}, in the Hermitian-Yukawa-interaction case, upon the replacement of the UV cutoff $\Lambda$ in the effective theory of that work with the transmutation mass $\tilde \mu$ in our approach. However, there is an \emph{essential difference} in our case from that of \cite{R15}, in that there is a nontrivial self interaction, which is necessarily nonvanishing, and its coupling is proportional to the negative of the square of the Yukawa coupling~\eqref{e46}. The couplings are both in the Hermitian regime. Any possible non-Hermitian regime in this analysis would lead to the generation of very large masses, which is physically unacceptable in perturbation theory.

%%%%%%%%%%%%%%%%%%%%%%%%%%% 
  \item Let us look for solutions with $b \ll a^{2}$. We deduce from \eq{e43c} that
  \begin{equation}
\label{e47}
\log a^{2}=\left(-\frac{4}{\overline{g}^{2}}+1-\gamma\right)-\frac{b}{a^{2}}\left(1-\gamma-\frac{4}{\overline{g}^{2}}\right)+\frac{b\log b}{a^{2}}
\end{equation}
and from \eq{e43b} that
\begin{equation}
\label{e48}
\overline{u}b\log b=-b\left(1+\overline{u}\left(\gamma-1\right)\right)-\overline{g}^{2}a^{2}\left(\gamma-1+\log a^{2}\right)
\end{equation}
For $b \ll a^{2}$ we have approximately from \eqref{e47} that
\begin{equation}
\label{e49}
a^{2}\simeq\exp\left(1-\gamma-\frac{4}{\bar{g}^{2}}\right).
\end{equation}
Hence 
\begin{equation}
\label{e49a}
\bar{u}b\log b=-\left(\bar{u}\left(\gamma-1\right)+1\right)b+4a^{2}.
\end{equation}
In order to be compatible with $b \ll a^{2}$, dominant balance requires
\begin{equation}
\label{e49b}
\bar{u}b\log b\simeq4a^{2}
\end{equation}
and also
\begin{equation}
\label{e49c}
\left|\bar{u}\log b\right|\gg\left|\bar{u}\left(1-\gamma\right)-1\right|.
\end{equation}
Since $a^2$ is positive and $b$ is small, from \eqref{e49b} we have $\bar{u}<0$ and so
\begin{equation}
\label{e49d}
-\log b\gg1-\gamma+\frac{1}{\left|\bar{u}\right|}.
\end{equation}
From \eqref{e49} and \eqref{e49b} we can show that for Hermitian $\bar{u}$ and $\bar{g}$ there is a possibility of generating masses in this regime when $\left|\bar{u}\right|$ is an order of magnitude smaller than $\bar{g}$.

    \item  Let us look for solutions with $b \gg a^{2}$. This case includes the possibility of zero fermion mass as well. From \eqref{e43bcc} we have on invoking dominant balance
    \begin{equation}
\label{e50}
b\approx\exp\left[1-\gamma-\frac{1}{\bar{u}}\right].
\end{equation}
For $b$ to be small we require small positive  $\bar{u}$, i.e. $u$ is non-Hermitian.
From \eq{e43c} we have 
\begin{equation}
\label{e50a}
a^{2}=b\left[\frac{4-\frac{\bar{g}^{2}}{u}}{4+\gamma\bar{g}^{2}}\right]
\end{equation}

Consequently we require 
\begin{equation}
\label{e51}
\overline{u}\approx\frac{\overline{g}^{2}}{4}.
\end{equation}
This indicates that solutions with $b \gg a^{2}$ may be viable when $\bar{u}$ and $\overline{g}^{2}$ have the same positive sign. Hence in this mass regime we have a Hermitian Yukawa coupling case, $g^2 > 0$, and an anti-Hermitian scalar self-interaction. In this case, we deduce that both axion and fermion dynamically-generated masses are extremely suppressed, for perturbatively small real Yukawa couplings $g$.  As already mentioned, zero fermion masses are also compatible with this scenario. For anti-Hermitian Yukawa  couplings, as considered in \cite{R15a,R11,R15},   $g^2 < 0$ (i.e. purely imaginary $g = i \breve g$, $\breve g \in \mathbb R$), this case leads to very large axion masses for $|\breve g| \ll 1$.  We stress once again that our results above point to an essential difference from the studies in \cite{R15a,R11,R15}, where self-interactions of the axions were not considered; here the pseudoscalar self-interaction coupling $u$ is necessarily nontrivial for consistency of the quantum theory. 

\end{itemize}

\section{Two-loop renormalisation group analysis: Renormalization Group Flows between Hermitian and non-Hermitian fixed points\label{sec:2loop}}

The presence of both Hermitian and non-Hermitian fixed points within our models might be the result of the one loop nature of our approximation. It is of  course, in general, difficult to rule out this possibility without some parameter in the theory which can control the contributions of higher loops.  However we have analysed a two loop renormalisation flow~\cite{R18,R19} for a similar, but massless, Yukawa model given by the Lagrangian $\mathcal{L}_{MY}$. In what follows, we shall demonstrate that in such a model, there is a renormalisation-group flow between Hermitian and non-Hermitian fixed points. 

 The Lagrangian $\mathcal{L}_{MY}$ is:

\begin{equation}
\label{e44a }
\:\mathcal{L}_{MY}=\frac{1}{2}\left(\partial\phi\right)^{2}+i\bar{\psi}\gamma^{\mu}\partial_{\mu}\psi+i\,g\,\phi\bar{\psi}\gamma_{5}\psi -\frac{u}{4!}\phi^{4}
\end{equation}
where, $\psi$ is a massless Dirac-fermion field and $\phi$ is a massless pseudoscalar field, $g$ denotes the Yukawa coupling,  and $u$ denotes the self-interaction of $\phi$. We shall consider $u>0$ (the Hermitian case) but allow $g$ to be real or imaginary. From the consideration of the convergence of path integrals given earlier we know that the usual Feynman rules are valid. If $u$ were to go towards a negative $u$ fixed point, according to the renormalisation group flow, then for a resulting $\left| u\right|  $ which is not small might be indicative of an interesting behaviour. If $\left| u\right|  $ is small then the Feynman rules would be still valid  since the Feynman rules give an approximation to the behaviour near the trivial saddle point of the path integral.

We define for notational convenience 
\begin{align}\label{newdefgh}
\tilde g \equiv \frac{g^2}{16\pi^2} \quad  {\rm and} \quad   
h \equiv \frac{u}{16\pi^2}~.
\end{align}
The loop calculation involves 14 topologically distinct graphs. In \cite{R18, R19} the calculation of the beta function $\beta_{\tilde g}$ for $\tilde g$ gives~\footnote{In $D=4-\epsilon$ the beta functions for the couplings in the model would have $\epsilon$ dependent terms determined by the engineering dimensions of the couplings in the noninteger $D$ dimensions.} 
\begin{equation}
\label{e45a}
\beta_{\tilde g}=10\, \tilde g^{2}+\frac{1}{6}\, h^{2} \, \tilde g - 4\, h\, \tilde g^{2}-\frac{57}{2}\, \tilde g^{3}
\end{equation}
and the calculation of the  beta function $\beta_{h}$ for $h$ gives
\begin{equation}
\label{e46a}
\beta_{h}=3\, h^{2} + 8\, h\, \tilde g - 48\, \tilde g^{2}-\frac{17}{3}\,h^{3}-12\, \tilde g\, h^{2}+28\, h\, \tilde g^{2}+384\, \tilde g^{3}.
\end{equation}

We can show that there are four fixed points $(\tilde g_{i},h_{i}), i=1,\cdots,4$ where
\begin{eqnarray}
\tilde g_{1} & = & 0 \, \quad \qquad \qquad \qquad h_{1} = 0\label{e51a}  \\
\tilde g_{2} & = &0 \, \quad \qquad \qquad \qquad h_{2} = 0.529412\label{e51b}   \\
\tilde g_{3} & = & -0.00570795 \, \qquad h_{3} = 0.525424\label{e51c}   \\
 \tilde g_{4}& = & 0.234024 \qquad  \qquad h_{4} =  1.01657 \label{e52}
\end{eqnarray}

In this two loop  calculation we note the appearance of a non-Hermitian (purely imaginary ({\it cf.} \eqref{newdefgh}) Yukawa coupling $g$ at  the $i=3$ fixed point.

\subsubsection{Stability analysis\label{sec:analysis}}

A linear stability analysis at the fixed point $({\tilde g_{i}}^{*},{h_{i}}^{*})$ in the $(\tilde g,h)$ coupling space gives
\begin{enumerate}
  \item for $i=2$  the eigenvalues $-1.58824, \quad 0.0467128$
  \item for $i=3$  the eigenvalues $-1.51242, \quad -0.0479488$
  \item for $i=4$  the eigenvalues $-13.1652, \quad-2.34048$.
  \end{enumerate}
None of these fixed points are IR stable (where all the eigenvalues are positive). 

The trivial fixed point $i=1$ requires a separate nonlinear analysis. The flow near the trivial fixed point is approximated by
\begin{eqnarray}
\frac{d \tilde g}{dt} & = & 10\, \tilde g^{2} \label{e53}\\
\frac{dh}{dt} & = & 3h^{2}+8h\, \tilde g-48\, \tilde g^{2} \label{e54}
\end{eqnarray}
The solution of (\ref{e53}) (for $\tilde g(t)$) is 
\begin{equation}
\label{e55}
\tilde g(t)=\frac{\tilde g_{0}}{1-10\, \tilde g_{0}t}
\end{equation}
with $\tilde g(0)=\tilde g_{0}$. For $\tilde g_{0}>0$  (the  Hermitian case) the flow is away from $\tilde g=0$ in the UV ($t \to \infty$); in fact, we observe from \eqref{e55}, that in the UV limit, the coupling $\tilde g$ approaches $0$, but from negative values, that is, there is a {\it flow} from Hermitian to non-Hermitian Yukawa couplings.  The point $\tilde g=0$ is IR ($ t \to 0$)  stable though.
For the \emph{non-Hermitian} case, $\tilde g_{0}<0$, the renormalization group flow in the UV is towards $\tilde g=0$. The Yukawa coupling stays non-Hermitian during the flow. In the IR, the renormalisation flow is away from $\tilde g=0$.

 Let us consider the behaviour of $h(t)$ with $h(0)\equiv h_{0}$; the solution of (\ref{e54}) is 
 \begin{equation}
\label{e56}
h\left(t\right)=\tilde g_{0}\frac{11.0416+13.0416\left(1-10\, \tilde g_{0}t\right)^{2.40832}\, c}{3\left(1-10\, \tilde g_{0}t\right)\left[1+\left(1-10\, \tilde g_{0}t\right)^{2.40832}\, c\right]}
\end{equation}
where $c=\frac{3.68053\, \tilde g_{0}-h_{0}}{-4.3472\, \tilde g_{0}+h_{0}}$. (The sign of $c$ is not important for the stability analysis.) For the \emph{Hermitian} case, $\tilde g_{0}>0$, the renormalisation group leads to a flow away from $h=0$ towards a non-Hermitian value of $h$ in the UV  and a flow towards $h=0$ in the IR. For the \emph{non-Hermitian} case, $\tilde g_{0}<0$, $h(t)$  flows to $0$ in the UV and in the IR $h(t)$ flows away from $h=0$ through Hermitian values of $h$. Hence we see an interplay of Hermitian and non-Hermtian behaviour in the renormalization group behaviour.\footnote{It would interesting to compare the non-Hermiticity in our model with spontaneous non-Hermiticity discussed in \cite{chern} within the traditional NJL model. } 

The possible connection between Hermitian and non-Hermitian fixed points that we have noticed is unlikely to be an artefact. There is some independent evidence that this happens in other theories although the possible connection with \cPT symmetry  was not realised. This independent evidence has been found  in a more complicated model, a chiral Yukawa model~\cite{R20a}, with the Standard Model symmetry implemented only at the global level.The flow of the quartic scalar coupling from positive to negative values was observed. However whether a \cPT interpretation is valid in detail remains to be explored.

\section{Conclusions and Outlook \label{sec:concl}}

In this work we have laid the foundation for the analysis of field theories involving \cPT symmetric interactions between a pseudoscalar and a Dirac fermion using a path integral formulation..
We have studied a perturbative renormalization-group analysis of the model, given in  \eqref{ee1}, involving a self-interacting pseudoscalar (axion-like) field coupled to fermions. The model, without axion self-interactions (and axion potentials),  has previously been considered from the point of view of dynamical mass generation within a Schwinger-Dyson framework~\cite{R11,R15}. We have noted here that quantum consistency at the perturbative level requires the presence of a non-trivial quartic  $\phi^4$ coupling, which is proportional to the square of the Yukawa interaction $g^2$, to leading order in perturbation theory in $g$.

Motivated by the models of \cite{R11,R15}, we have considered both Hermitian ($g^2 >0$) and anti-Hermitian ($g^2 < 0$) Yukawa couplings. As a preparation for a full Schwinger-Dyson treatment, which we postpone to a future publication,  we have studied here the possibilities for a one-loop dynamical mass generation for both axions and fermions using a method due to Nambu and Joan-Lasinio.

 We have focused here on a one-loop dynamical mass generation for both axion and fermion fields in the model  \eqref{ee1}, which we have studied in various limits. 
 Depending on the sign of the squared coefficient, one can obtain  Hermitian or anti-Hermitian axion self-interactions. In the Hermitian Yukawa interaction case, for the case of equal masses of axions and fermions, we have recovered the non-perturbative expression for the  dynamically generated masses of axions and fermions, discussed in \cite{R11,R15} (but in the presence of the axion self-interaction quartic coupling $u$).
 
We have also managed to demonstrate that there is a renormalization-group flow between Hermitian and non-Hermitian fixed points in the theory, which also manifests itself at two loops in a model with no bare axion and fermion masses. This last result might be interpreted as a spontaneous appearance of non-Hermiticity in this class of models, in analogy to the situation characterising Nambu-Jona-Lasinio theories with four-fermion interactions. The reader should also recall that, as far as the quartic (pseudoscalar) coupling $u$ is concerned, there is a flow from the Hermitian 
case ($u <0$), corresponding to a positive $\phi^4$ potential ({\it cf.} \eqref{e1}), to 
the non-Hermitian (\cPT symmetric) 
case ($ u > 0)$ corresponds to an upside down $\phi^4$ self-interaction potential. 

\section*{Acknowledgments}

The work of N.E.M. and S.S. is supported in part by the UK Science and Technology Facilities research
Council (STFC) and UK Engineering and Physical Sciences Research
Council (EPSRC) under the research grants ST/T000759/1 and  EP/V002821/1, respectively. NEM acknowledges participation in the COST Association Action CA18108 {\it Quantum Gravity Phenomenology in the Multimessenger Approach (QG-MM)}. We would like to thank Carl Bender and Wen-Yuan Ai for valuable discussions.
\vspace{.5cm}

\appendix{}

\section{Aspects of Pseudohermiticity\label{sec:appA}}
\vspace{.2cm}
It was pointed out by Mostafazadeh~\cite{R3} that \cPT symmetry was part of a more general framework known as pseudoHermiticity. We know that \cPT symmetry implies 
\begin{equation}
\label{pseudoHerm }
H=\mathcal{P}^{-1}H^{\dag }\mathcal{P}=\mathcal{P}^{}H^{\dag }\mathcal{P}
\end{equation}
since $\mathcal{P}=\mathcal{P}^{\dag }$ and $\mathcal{P}^{2}=1$.

In quantum mechanics any operator  $H$ (acting on a Hilbert space $\mathcal{H}$) is pseudoHermitian if it can be related to its adjoint by
\begin{equation}
\label{pseudo}
H^{\dag }=\eta H \eta^{-1} 
\end{equation}
where $\eta$ is a bounded automorphism of the Hilbert space ($\eta$ can be chosen to be Hermitian). Pseudo-Hermiticity  is  a generalisation  of both Hermiticity and \cPT symmetry.
If the usual model-independent inner product on $\mathcal{H}$ is written as $\langle |\rangle $ then 
\begin{equation}
\label{inner}
\langle \phi |H\psi \rangle =\langle H^{\dag }\phi |\psi \rangle .
\end{equation}
 The pseudoHermitian  $H$ can have both real and complex conjugate eigenvalues. Let us consider a form on  $\mathcal{H}$ defined by
\begin{equation}
\label{inner2}
\langle \phi |\psi \rangle_{\eta} \equiv \langle \phi |\eta|\psi \rangle =\langle \phi |\eta\psi \rangle =\langle \eta\phi |\psi \rangle .
\end{equation}
The adjoint with respect to this inner product, $\widehat{H}$ say, is defined by 
\begin{align}
\label{inner3}
  \langle \widehat{H} \phi |\psi \rangle_{\eta}&=\langle \phi |H\psi \rangle_{\eta} =\langle \phi | \eta H\psi \rangle   \\
    &=  \langle H^{\dag }\eta\phi |\psi \rangle_{} =\langle \eta \eta^{-1}H^{\dag }\eta\phi |\psi \rangle_{} \\
    &= \langle \eta^{-1}H^{\dag }\eta\phi |\psi \rangle_{\eta} 
\end{align} 
So ${}_{}\widehat{H} =\eta^{-1}H^{\dag }\eta$ and 
\begin{equation} 
\label{inner4}
\widehat{H} =H.
\end{equation}
In order to have a probabilistic interpretation for the quantum mechanics  in terms of this  inner product, it is necessary to choose $\eta$ to be a positive operator $\eta=\tilde \eta^{\dag }\tilde\eta$ . We can write 
\begin{equation}
\label{inner5}
H=\tilde \eta^{-1}\left( \tilde \eta^{-1}\right)^{\dag }  H^{{}^{\dag }}\tilde \eta^{\dag }\tilde \eta
\end{equation}
and so
\begin{equation}
\label{inner6}
\tilde \eta H \tilde \eta^{-1}=\left( \tilde \eta^{-1}\right)^{\dag }  H^{{}^{\dag }}\tilde \eta^{\dag }=\left( \tilde \eta H \tilde \eta^{-1}\right)^{\dag }. 
\end{equation} 
If we identify 
\begin{equation}
\label{inner7}
h=\tilde \eta H \tilde \eta^{-1}
\end{equation}
 then form \eq{inner6}  we find
\begin{equation}
\label{inner7b}
h=h^{\dag}.
\end{equation}
Hence for any pseudoHermitian $H$ we can in principle find a corresponding  Hermitian $h$ using a similarity transformation $q$ which may not be unique. This argument would also be valid for other pseudoHermitian operators.This result has been found in a quantum system with 
 a finite number of degrees of freedom. For a field theory with an infinite number of degrees of freedom it may be possible to formally construct $\tilde \eta $ and requires further investigation ({\it cf.} Eq.\eq{simil}).
%%%%%%%%%%%%%%%%%%%%%%%%%%%%%%%%%%%%%%%%%

%%%%%%%%%%%%%%%%%%%
%%%%%%%%%%%%%%%%%%%

\end{document}